\documentclass[a4paper,12pt,dvipsnames]{article}

\usepackage[top=2cm, bottom=2cm, left=2.5cm, right=2.5cm]{geometry}
\usepackage{amsmath}
\usepackage{amssymb,amsfonts,amsthm}
\newcommand\numberthis{\addtocounter{equation}{1}\tag{\theequation}}
\newcommand{\indep}{\perp\!\!\!\perp}
\usepackage{bm}
\usepackage{hyperref}
\usepackage{setspace}
\usepackage[justification=centering]{caption}
\usepackage{float}
\usepackage{xcolor}
\usepackage{changes}
\usepackage{multirow}
\usepackage{longtable}
\usepackage{supertabular} 
\usepackage{dcolumn}
\usepackage{pgfplots}
\usepackage{pdflscape}
\usepackage{float}
\usepackage[shortlabels]{enumitem}
\usepackage{longtable}
\usepackage{tabularx,calc}
\usepackage{booktabs}
\usepackage{acro}
\usepackage{setspace}
\usepackage{afterpage}
\usepackage{lscape}
\usepackage{mathtools}
\usepackage{pst-plot, pstricks} 
\usepackage{fancybox,amssymb,color}
\usepackage{subcaption}
\usepackage{framed}
\usepackage{titling}
\usepackage[bottom]{footmisc} 
\usepackage[]{authblk}
\usepackage[title]{appendix}

\usepackage[backend=biber, style=apa, maxcitenames=2, sorting=nyt, sortcites=false]{biblatex}
\usepackage[most]{tcolorbox}

\definecolor{Diceblue}{RGB}{52,116,181}

\hypersetup{
	colorlinks = true,
	linkcolor=Diceblue,
	citecolor=Diceblue,
	urlcolor=Diceblue
}

\begin{document}

	
\setlength{\abovedisplayskip}{3pt}
\setlength{\belowdisplayskip}{12pt}
	
	
\setlength{\jot}{0.5cm}

\begin{titlingpage}

\title{Bayesian Machine Learning Methods For Large Scale Demand Estimation}

\author{
Anna B. Schmidt$^\ast$
}

\date{\small\today}

\maketitle

\begin{abstract}
		
This work studies how Bayesian machine learning methods can be used for large-scale demand estimation with many product categories. I compare two model classes, a latent factorization model and a mixed logit model and two Bayesian estimation approaches, Markov Chain Monte Carlo (MCMC) and Variational Inference (VI). The analysis combines a simulation study with an application to supermarket scanner data. The results show that the latent factorization model benefits from information across categories and improves its predictive performance as the dimensionality of the choice environment increases, whereas the mixed logit model does not exhibit the same pattern. MCMC delivers the highest predictive accuracy but is computationally intensive. VI achieves slightly lower predictive performance while substantially reducing runtime. In the empirical application, VI also outperforms the mixed logit benchmark. These findings highlight a trade-off between accuracy and computational feasibility in multi-category demand estimation.

		

\vspace{0.4cm}
		
 \vspace{0.1cm}\newline

\end{abstract}

\vfill
\noindent\hrulefill \\
{\footnotesize
I gratefully acknowledge financial support by the German Research Foundation (DFG, grant 462020252) and thank Florian Heiss and Daniel Brunner for their supervision and helpful comments. Computational infrastructure and support were provided by the Center for Information and Media Technology (ZIM) at Heinrich Heine University Düsseldorf. 
\begin{description}
    \setlength{\itemsep}{0pt}
    \setlength{\parskip}{0pt}
    \setlength{\parsep}{0pt}
    \item[$^\ast$] Heinrich Heine University Düsseldorf, Chair of Statistics and Econometrics, Universitätsstr. 1, 40225 Düsseldorf, Germany; e-mail: \texttt{\href{mailto:anna.brigitte.schmidt@hhu.de}{anna.brigitte.schmidt@hhu.de}}
\end{description}
}

\end{titlingpage}


\onehalfspacing

\newpage
\pagenumbering{arabic} 
\setcounter{page}{1}
\section{Introduction}

In the context of the global advancement of digitalization within commerce and industry, there has been a substantial increase in the volume of data generated on a daily basis. In 2023, the estimated volume of created, captured, copied and consumed data worldwide amounts to 328.77 quintillion bytes (equivalent to $328.77 \times 10^{18}$) or 0.33 zettabytes daily \parencite{Statista2023}. 

Therefore, scalable methods of processing such extensive data like machine learning methods have become very popular in nearly all scientific fields (for a comprehensive review of machine learning methods in economics see \textcite{Athey_2019MLreview}).
In the context of demand estimation, the availability of individual or household level scanner data have increased the interest in machine learning methods tremendously \parencite{Bajari2015MLdemand}. 

Due to computational constraints, early studies like \textcite{Chiang1991simultaneous} were limited to examining demand from discrete choice data in only a single category. A model focused solely on a single category offers an incomplete representation of consumer behaviour, overlooking potential interdependencies in purchasing decisions across various product categories. For example, the brand loyalty of a consumer might persist across multiple categories. In marketing, this concept is recognized as the signaling theory of umbrella branding \parencite{Erdem1998empirical}. \textcite{ErdemWiner1999} found strong empirical evidence for cross-category correlations within a given brand name across two product categories.

More recently, factorization models gained recognition for multi-category discrete choice estimation in work conducted by \textcite{Athey2018ttfm}, \textcite{Ruiz2020shopper}, \textcite{Donnelly2021counterfactual} and \textcite{wan2017modeling}. These models allow for rich heterogeneity in consumer preferences. They emerged in the recommender system literature and have become popular in this field since a matrix factorization based collaborative filtering algorithm won the Netflix price competition in 2009 for predicting movie ratings \parencite{koren2009matrix}.

\textcite{Donnelly2021counterfactual} apply the assumption that consumers purchase just one item from each category during a single shopping trip. Additionally, they introduce a nested logit model structure. This approach permits correlation of utilities across different products within the same category. Thereby, it provides a more accurate reflection of consumers' decision-making processes, including whether to buy from a particular category at all.
\textcite{Ruiz2020shopper} focus their work on identifying substitutes and complements by analysing shopping baskets.  
\textcite{wan2017modeling} put forward a distinct model for estimating consumer grocery demand, employing a latent factorization model structured in three stages. Initially, the consumer faces a binary decision for every category, determining whether to purchase an item from that category. Following this decision, the consumer selects which specific item to buy from the category, utilizing a multinomial choice process. In the final stage, the consumer decides the quantity of the chosen item to purchase, with the number drawn from a Poisson distribution.

The aim of this work is to explore potential enhancements in predictive accuracy across multiple categories. To achieve this, both a factorization model and a mixed logit model are employed for comparative analysis. Given that multi-category models often involve estimating a large number of parameters, they pose computational difficulties, especially when using frequentist methods \parencite{Seetharaman2005models}.

Therefore, this work will focus on Bayesian machine learning methods. 
In the Bayesian framework, the objective is to derive knowledge about a \textit{posterior} distribution $p(\theta|X)$, where $\theta$ is treated as a random parameter vector and $X$ is the observed data, by applying Bayes' theorem \parencite{Gelman_et_al_2014}:
\begin{align} \label{eq: Bayes}
	p(\theta|X) = \frac{p(X | \theta) \ p(\theta)}{p(X)}.
\end{align}

Here, the probability of observing data $X$ given parameters $\theta$ is called the \textit{ likelihood} $p(X|\theta) $ and $p(\theta)$ is the \textit{prior} distribution of the parameters.
The denominator is referred to as \textit{marginal likelihood} or \textit{evidence}, which can be defined as 
\begin{align}
	p(X) = \int p(X|\theta) p(\theta) d \theta.
\end{align}

The marginal likelihood involves integrating over all possible parameter values $\theta$. This integration can be high-dimensional and intractable, especially for complex models with a large amount of parameters like factorization models. This work will provide a detailed exploration of two prominent techniques for approximating the posterior distribution, specifically \textit{Markov Chain Monte Carlo} (MCMC) and \textit{Variational Inference} (VI).

The remainder of this paper is structured as follows. Section \ref{sec: methods} provides a comprehensive description of Bayesian machine learning methods, specifically focusing on MCMC and VI. This section illustrates commonly employed algorithms such as  \textit{Metropolis-Hastings} (MH), \textit{Hamiltonian Monte Carlo }(HMC), \textit{Coordinate Ascent Variational Inference }(CAVI) and \textit{Automatic Differentiation Variational Inference} (ADVI). 
Next, section \ref{sec: models} will focus on the discrete choice models used in this research, namely the mixed logit model and the factorization model. These models are examined in the framework of utility maximizing individuals. The underlying assumptions and mathematical formulations are explicated, providing a foundation for the subsequent analyses.
Section \ref{sec: study} outlines a simulation study, designed to validate the methods proposed in the preceding sections. This includes a detailed description of the \textit{data generating process} (DGP), computational specifications, results and a discussion of the findings from the simulation.
Afterwards, section \ref{sec: real_data} extends the simulated analyses to a real-world context by applying the developed methods to a dataset of supermarket scanner data. This section encapsulates data characteristics, empirical results and a critical discussion.
Concluding the thesis, section \ref{sec: conclusion} offers a summary of the principal findings, practical implications and prospective avenues for future research.

The major findings are that the latent factorization model gains predictive accuracy from observing an increasing number of variables, while the mixed logit model does not exhibit this behaviour. While MCMC provides the highest prediction accuracy, it comes with substantial computational overhead. On the other hand, VI is the most time-efficient approach with only a slight reduction in predictive accuracy.
In the real data application, VI outperforms the mixed logit model in terms of prediction accuracy and runtime. At the current state of implementation, MCMC seems to be infeasible for the large scale dataset, due to its intensive computational requirements.

\section{Methods} \label{sec: methods}

Computation of the posterior is often intractable. Therefore, approximation methods are needed to perform Bayesian inference. The most popular approximation techniques are MCMC and VI, which are explained in more detail in the following chapter.
\subsection{Markov Chain Monte Carlo}

MCMC is a prevalent method for Bayesian inference which aims to approximate the posterior distribution $p(\theta | X)$ through sampling from the exact posterior \parencite{Gelman_et_al_2014}.  \textcite{metropolis1953equation} first applied MCMC to simulate the distribution of states for a system of idealized molecules. As the name implies, this method is based on two components: Markov chains and Monte Carlo simulation. Markov chains possess some features that make them suitable for sampling from the desired posterior distribution. These characteristics will be explained in more detail in the following.

Let $(\Omega, \mathcal{F}, p)$ be a probability space with sampling space $\Omega$, $\sigma$-algebra $\mathcal{F}$ on $\Omega$ and probability function $p$.
A sequence of random variables $(X_n)_{n \in \mathbb{N}_0}$ on $(\Omega, \mathcal{F}, p)$ is called a Markov chain, if the distribution of $X_{n + 1}$ is conditionally independent of all preceding $(X_m)_{m < n}$ given $X_{n}$ \parencite{Neal1993probabilistic}:
\begin{align} \label{eq: markov}
	X_{n + 1} \indep (X_m)_{m < n} \ | \ X_{n}.
\end{align}
The range of these $X_n$ is called the \emph{state space} $\mathcal{S}$ of the Markov chain and every element of $\mathcal{S}$ is called a \emph{state}. For a continuous state space the probability of transitioning from state $x \in \mathcal{S}$ to a state in the measurable set $A \subset \mathcal{S}$ is defined by the transition kernel $P(x, A)$ with $P(x, \mathcal{S}) = 1$ \parencite{Tierney_1994_MCMC}. 
Therefore,  the $n$-th iterate of the kernel starting from $x$ is denoted as
\begin{align}
	P^{(n)}(x, A) = \int_{\mathcal{S}} P^{(n-1)}(x, dy) \ P(y, A),
\end{align}
where $P^{(n-1)}(x, dy)$ represents the probability of transitioning from $x$ to a new state $y$ in exactly $n - 1$ steps and $dy$ indicates that the integral is taken over all possible intermediate states represented by $y$. $P(y, A)$ is the probability of transitioning from state $y$ to a state in set $A$ in a single step.

A Markov chain is said to have a \textit{stationary} or \textit{invariant} distribution $\pi(\cdot)$ if this distribution persists forever once it is reached
\begin{align}
	\pi(A) = \int_{\mathcal{S}} P(x, A) \ \pi(dx). 
\end{align}
The goal of MCMC is to construct a Markov chain, whose unique stationary distribution $\pi(\cdot)$ corresponds to the target posterior distribution $p(\theta | X)$. Therefore, the long term behaviour $n \rightarrow \infty$ of a chain is of particular interest. 
\textit{Ergodicity} describes the asymptotic convergence of a Markov chain towards its invariant distribution as the number of iterations approaches infinity \parencite{RobertCasella2004}. Under suitable regularity conditions, this convergence holds independently of the initial state.
A Markov chain is ergodic if it is irreducible, aperiodic and has $\pi(\cdot)$ as its stationary distribution. 
For a continuous state space, a Markov chain is \textit{irreducible} if every measurable set of positive reference measure can be reached from any initial state with positive probability in a finite number of steps. \textit{Aperiodicity} means that the chain does not follow a fixed cyclic pattern. Assuming $x, y \in \mathcal{S}$, a Markov chain is \textit{reversible} if the following condition is satisfied:
\begin{align} \label{eq: reversible}
	\pi(x) \ p(x, y) = \pi(y) \ p(y, x).
\end{align}
Here, $p(x, y)$ and $p(y, x)$ are functions describing the chain moving from state $x$ to $y$ and vice versa. This property is also called \textit{detailed balance} because it can be understood as a balance between the probabilities of moving forward and backward between states \parencite{ChibGreenberg1995MH}. Detailed balance provides a sufficient, but not necessary, condition for invariance.

Since the transition kernel is unknown in MCMC methods, various algorithms such as MH or HMC can be employed to construct an appropriate transition kernel that satisfies the required conditions.
Detailed explanations of these algorithms will be provided in the subsequent chapters.

\subsubsection{Metropolis-Hastings Algorithm}
The MH algorithm was developed by \textcite{hastings1970monte} based on the earlier work of \textcite{metropolis1953equation}. The following explanation of this algorithm is drawn from the work by \textcite{ChibGreenberg1995MH}. 

They showed that the reversibility condition in equation \eqref{eq: reversible} is a sufficient condition for $\pi(\cdot)$ being the stationary distribution. First, a proposal density $q(x,y)$ with $\int q(x, y) dy = 1$ is needed to generate a new value $y$ from the current state $x$. The proposal functions $q(x, y)$ and $q(y, x)$ describe the probability that the chain moves from state $x$ to $y$ and vice versa. Since the functional form of $q(x, y)$ is not explicitly specified, there is a possibility that the process might frequently move from state $x$ to state $y$, while rarely transitioning in the opposite direction
\begin{align} \label{eq: ineq}
	\pi(x) \ q(x, y) > \pi(y) \ q(y, x).
\end{align}
To address this issue, one can introduce an acceptance probability denoted as $\alpha(x, y) < 1$, which reduces the probability of transitioning from state $x$ to state $y$. At the same time, $\alpha(y, x)$ is set to 1, ensuring that transitions from state $y$ to state $x$ are always accepted. Then reversibility is satisfied if
\begin{align}
	\pi(x) \ q(x, y) \ \alpha(x, y) &= \pi(y) \ q(y, x) \ \alpha(y, x) \\
	&=  \pi(y) \ q(y, x) \\
	\iff  \alpha(x, y) &= \frac{\pi(y) \ q(y, x)}{\pi(x) \ q(x, y)}. 
\end{align}
If the inequality in \eqref{eq: ineq} is reversed, set $\alpha(x, y) = 1$ and obtain $\alpha(y, x)$ analogously.
This leads to an acceptance probability of 
\begin{align} \label{eq: acc_prob}
	\alpha(x, y) = \begin{cases}
		\text{min} \Big(1, \frac{\pi(y) \ q(y, x)}{\pi(x) \ q(x, y)} \Big) & \text{if} \  \pi(x) \ q(x, y) > 0 \\
		1 & \text{otherwise.}
	\end{cases}
\end{align}
Consequently, the probability to reject $y$ and stay at the current state $x$ is $r(x)= 1 - \int_{\mathcal{S}} q(x, y) \ \alpha(x,y) dy$.
The appropriate transition kernel of the MH algorithm is
\begin{align} \label{eq: kernel_MH}
	P_{MH}(x, dy) = q(x, y) \ \alpha(x, y) \ dy + r(x) \ \delta_x(dy),
\end{align}
where $\delta_x(dy)$ is the Dirac measure, which equals to 1 if $x \in dy$ and 0 otherwise.

Finally, the MH algorithm can be summarized in the following steps to generate $N$ samples:
\begin{enumerate}
	\item Draw initial sample $x^{(0)}$.
	\item Repeat for $t = 1, \dots, N$:
			
			Draw $y \sim q(x^{(t)}, y)$ and  $u \sim \mathcal{U}(0, 1)$. 
			
			Calculate $\alpha(x^{(t)}, y)$ from equation \eqref{eq: acc_prob} and set $x^{(t + 1)} = y$ if $u \leq \alpha(x^{(t)}, y)$. Otherwise, set $x^{(t + 1)} = x^{(t)}$.
\end{enumerate}

Because the acceptance probability only depends on the current state $x$, the generated samples satisfy the Markov property from equation \eqref{eq: markov}. In addition, they are aperiodic and irreducible, as long as the proposal density $q(x, y)$ has a positive density on the support of $\pi(\cdot)$. Since they are aperiodic, irreducible and reversible, the kernel in equation \eqref{eq: kernel_MH} constructs an ergodic Markov chain whose stationary distribution is the desired target distribution independent from the initial sample $x^{(0)}$.

\begin{figure}[H] 
	\begin{center}
		\vspace{-3ex}
		\includegraphics[width=1.4\textwidth]{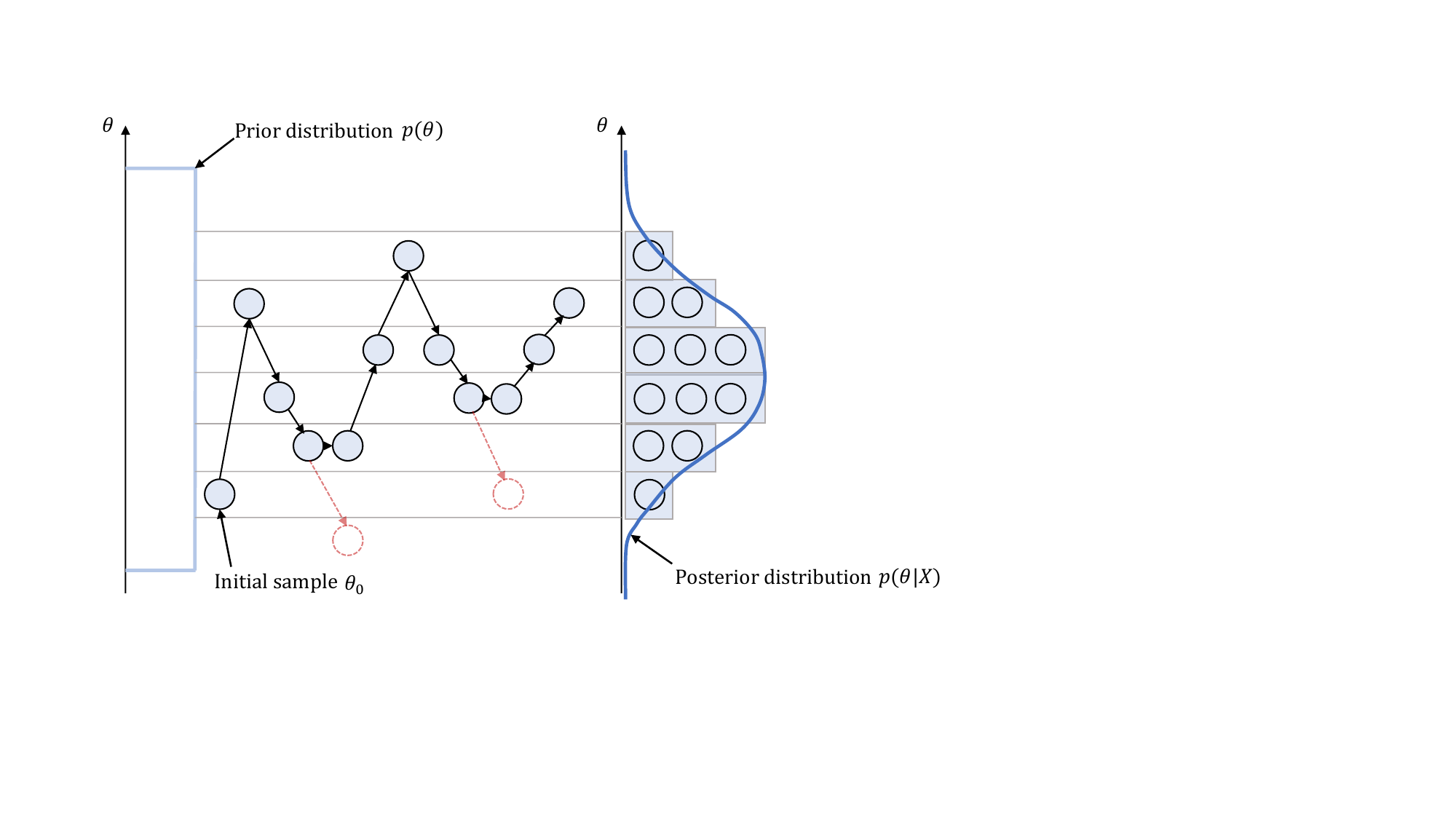}  
	\end{center}
	\vspace{-10ex}
	\caption{Schematic illustration of the MH algorithm with a uniform prior distribution and a normal posterior distribution. Depicted corresponding to \textcite{Lee_et_al_MH_pic}. }
	\label{abb: MH}	
\end{figure}

Figure \ref{abb: MH} shows a schematic illustration of this algorithm for a uniform distributed prior and a normal distributed posterior, where the red dots refer to rejected proposals. In this Bayesian setting the target distribution is the posterior $p(\theta | X)$ and therefore the acceptance probability from equation \eqref{eq: acc_prob} of transitioning from $\theta$ to $\theta'$  becomes
\begin{align}
	\alpha(\theta, \theta') = \begin{cases}
		\text{min} \Big(1, \frac{p(X| \theta') p(\theta') \ q(\theta', \theta)}{p(X|\theta) p(\theta) \ q(\theta, \theta')} \Big) & \text{if} \  p(X|\theta) p(\theta) \ q(\theta, \theta') > 0 \\
		1 & \text{otherwise,}
	\end{cases}
\end{align}
where $p(X|\theta)$ is the likelihood and $p(\theta)$ the prior.
An advantage of this algorithm is that it does not require the marginal likelihood $p(X)$ since it cancels out in the ratio above.

The number of iterations required for the algorithm to converge to the target distribution heavily depends on the choice of the proposal distribution. The phase before convergence is reached is commonly referred to as the \textit{warm-up} phase. 
In practice, a multivariate normal distribution is frequently used  to generate proposals \parencite{Betancourt2018conceptual}. However, \textcite{Roberts1997weakMH} demonstrate that, for higher dimensional target distributions, the optimal scaling of the proposal distribution results in an acceptance rate of only approximately $23.4\%$ of proposals. This relatively low acceptance rate reflects a trade-off between the size of proposed moves and their probability of acceptance. If the proposal variance is too large, a significant portion of proposals is rejected, whereas a proposal variance that is too small results in only small movements, both leading to a slow exploration of the target distribution by the Markov chain.

\subsubsection{Hamiltonian Monte Carlo} \label{sec: HMC}

The HMC algorithm was developed by \textcite{duane1987hybrid}, who referred to it as \textit{hybrid Monte Carlo}. It overcomes the inefficiency of the MH algorithm by using Hamiltonian dynamics to move more efficiently through the parameter space.

The intuition behind this approach is that one can view the statistical problem of moving efficiently through the parameter space of $\theta$ without being drawn towards the mode or tails of the posterior as a physical problem \parencite{Betancourt2018conceptual}. Consider a physical system involving a planet, a satellite and gravitational forces. In this analogy, the objective is to ensure that the satellite remains on the planet's orbit, steering clear of collision and preventing drifting into outer space. In physics, this is achieved by precisely applying an appropriate level of \textit{momentum} to the satellite.

\begin{figure}[H]
	\centering
	\begin{subfigure}{0.5\textwidth}
		\centering
			\begin{tikzpicture}
			\begin{axis}[
				axis x line=center,
				axis y line=none,
				xlabel={$\theta$},
				xmin=-3.5, xmax=3.5,
				ymin=0, ymax=0.5,
				samples=200,
				domain=-3.5:3.5,
				smooth,
				every axis plot/.append style={color=blue},
				legend style={at={(1,0.89)},anchor=north east}
				]
				\addplot[thick] {exp(-x^2 / 2) / (sqrt(2 * pi))};
				\addlegendentry{\(p(\theta|X)\)}
			\end{axis}
		\end{tikzpicture}
	\end{subfigure}%
	\begin{subfigure}{0.5\textwidth}
		\centering
		\begin{tikzpicture}
			\begin{axis}[ 
				axis x line=center,
				axis y line=center,
				axis line style = thick,
				xlabel={$\theta$},
				ylabel={$\phi$},
				xmin=-3.5, xmax=3.5,
				ymin=-3.5, ymax=3.5,
				samples=50,
				domain=-3.5:3.5,
				y domain=-3.5:3.5,
				view={0}{90},
				colormap/cool,
				legend style={at={(1,1)},anchor=north east}
				]
				\addplot3[
				mesh,
				colormap/cool,
				opacity=0.5
				] 
				{exp(-x^2 - y^2) / (2 * pi)};
				\addlegendentry{\(p(\phi, \theta)\)}
			\end{axis}
		\end{tikzpicture}
	\end{subfigure}
	\vspace{1.5ex}
	\caption{Transformation from parameter space $\theta$ (left) to phase space $(\phi, \theta)$ (right) for a normal posterior distribution.}
	\label{abb: HMC_trans}
\end{figure}
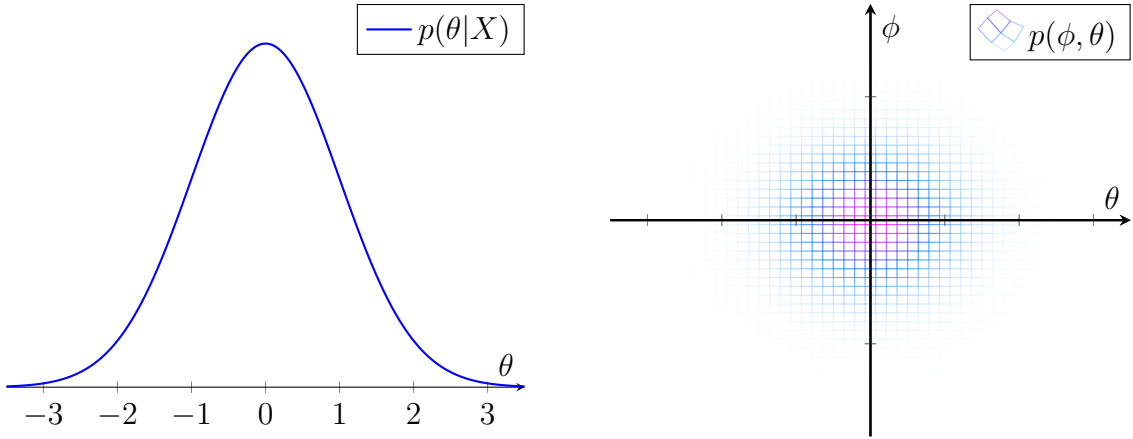

Therefore, the main idea of HMC is to augment the parameter space $\theta$ with an auxiliary momentum variable $\phi$. This augmentation creates the so called \textit{phase space} $(\phi, \theta)$. Figure \ref{abb: HMC_trans} provides an illustrative example of this transformation using a posterior distribution that follows a normal distribution. In addition, the transformation of the probability distribution $p(\theta | X)$ into the phase space is required.
The joint probability of the phase space is called the \textit{canonical distribution} and is given by the conditional distribution
\begin{align} \label{eq: canonical}
	p(\phi, \theta) = p(\phi  |  \theta)p(\theta).
\end{align}
To maintain clarity, the posterior $p(\theta|X)$ is represented as $p(\theta)$.

The canonical distribution is a concept from statistical mechanics and provides a connection between probabilities and energy functions, such as the Hamiltonian function $H(\phi, \theta)$ \parencite{Neal2011_HMC}. 
As a result, equation \eqref{eq: canonical} can be described as follows:
\begin{align}
	p(\phi, \theta) = e^{-H(\phi, \theta)}.
\end{align}
Conversely, the Hamiltonian function can be defined as
\begin{align}
	H(\phi, \theta) &= - log\big(p(\phi|\theta)p(\theta)\big) \\
	&= - log \ p(\phi|\theta) - log \ p(\theta)\\
	&= K(\phi, \theta) + V(\theta),
\end{align} 
where $K(\phi, \theta)$ is called the \textit{kinetic energy} and $V(\theta)$ is the \textit{potential energy} corresponding to the target posterior distribution. The kinetic energy is a tuning parameter of the algorithm and can be specified in various manners, such as an Euclidean-Gaussian, Riemann-Gaussian, or non-Gaussian kinetic energy. According to \textcite{Betancourt2018conceptual}, the prevailing choice is an Euclidean-Gaussian kinetic energy
\begin{align}
	K(\phi, \theta) = \frac{1}{2} \phi^T \cdot M^{-1} \cdot \phi + log \ |M| + \text{const.},
\end{align}
where $M$ represents a symmetric, positive-definite \textit{mass matrix}, which is usually diagonal.
Following the conversion to the phase space, it becomes possible to establish trajectories that maintain the imagined satellite in its orbit for some time $t$ through the utilization of Hamilton's equations:
\begin{align}
	\frac{d\theta}{dt} &= \frac{\partial H(\phi | \theta)}{\partial  \ \phi} = \frac{\partial K(\phi, \theta)}{\partial \  \phi} = M^{-1} \phi \\
	\frac{d \phi}{dt} &= - \frac{\partial H(\phi | \theta)}{\partial \ \theta} = - \frac{\partial K(\phi, \theta)}{\partial \ \theta} - \frac{\partial V(\theta)}{\partial \ \theta} = \frac{\partial log \ p(\theta|X)}{\partial \ \theta}.
\end{align}

However, these differential equations evolve in continuous time. Therefore, one needs to discretize the time with some small stepsize $dt \approx \epsilon > 0$. A suitable method for this discretization is called the \textit{leapfrog integrator} \parencite{Neal2011_HMC}. It consists of the following three steps to transition from $(\phi_t, \theta_t)$ to $(\phi_{t + \epsilon}, \theta_{t + \epsilon})$:

Fist, $\phi_t$ is updated a half-step 
\begin{align}\label{eq: leapfrog_1}
	 \phi_{t + \frac{\epsilon}{2}} &= \phi_t + \frac{1}{2}\epsilon \ \frac{\partial \ log(p(\theta_t |X))}{\partial \ \theta}.
\end{align}

Then the parameter vector $\theta_t$ makes on full step of length $\epsilon$ based on the half-updated $\phi_{t + \frac{\epsilon}{2}}$
\begin{align}\label{eq: leapfrog_2}
	\theta_{t + \epsilon} &= \theta_t + \epsilon M^{-1} \phi_{t + \frac{\epsilon}{2}} .
\end{align}

Lastly, $\phi_{t + \frac{\epsilon}{2}}$ is once again updated by half a step
\begin{align}\label{eq: leapfrog_3}
	 \phi_{t + \epsilon} &= \phi_{t + \frac{\epsilon}{2}} + \frac{1}{2}\epsilon \ \frac{\partial \ log(p(\theta_{t + \epsilon}|X))}{\partial \ \theta}.
\end{align}
These steps are repeated $L$ times to generate a trajectory of length $L$. The leapfrog integrator effectively simulates the continuous  Hamiltonian dynamics in discrete steps. It is a reversible and symplectic method, which means that it preserves phase-space volume and maintains the overall geometric structure of the phase space, but does not exactly conserve the Hamiltonian \parencite{Neal2011_HMC}. This is important for maintaining the quality of the sampled parameters, while the Metropolis acceptance step corrects for numerical error in the Hamiltonian.

\begin{figure}[H] 
	\begin{center}
		\vspace{-2ex}
		\includegraphics[width=0.7\textwidth]{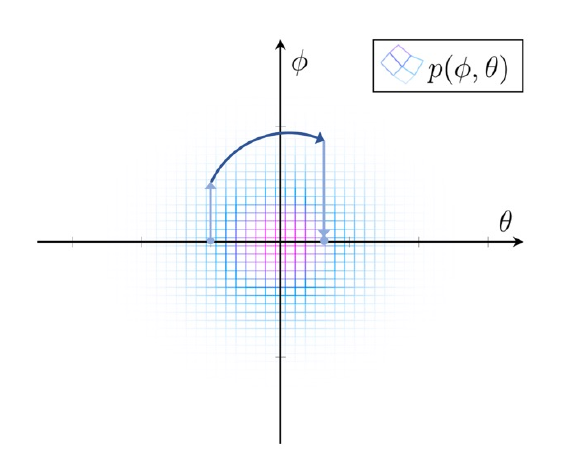}  
	\end{center}
	
	\caption{Representation of an iteration of the HMC algorithm in reference to the depiction in \textcite{Betancourt2018conceptual}. }
	\label{abb: HMC_iter}	
\end{figure}

Figure \ref{abb: HMC_iter} presents a visualization of this procedure to generate a single Hamiltonian Markov transition. This process involves a stochastic elevation from the target parameter space to the phase space (light blue), a deterministic Hamiltonian trajectory across the phase space (dark blue), and a subsequent projection back to the original target parameter space (light blue). Formally, one iteration of the HMC algorithm  is composed of the following steps \parencite{Gelman_et_al_2014}:
\begin{enumerate}
	\item Update $\phi \sim \mathcal{N}(0, M)$.
	\item Perform $L$ leapfrog steps, by calculating equations \eqref{eq: leapfrog_1}-\eqref{eq: leapfrog_3} for each step. 
	
	\item Label vectors at the start of the leapfrog process as $\theta^{t-1}, \phi^{t-1}$ and after the $L$ steps as $\theta^*, \phi^*$. Compute:
	\begin{align}
		\alpha = \frac{p(\theta^* | X)p(\phi^*)}{p(\theta^{t-1})p(\phi^{t-1})}.
	\end{align}
	
	\item Set \begin{align}
		\theta^t = \begin{cases}
			\theta^* & \text{with probability min}(\alpha, 1)\\
			\theta^{t-1} & \text{otherwise.}
		\end{cases}
	\end{align}
\end{enumerate}

Besides from the chosen kinetic energy, the efficiency of the HMC algorithm highly depends on the chosen step size $\epsilon$ and the number of steps $L$ \parencite{Neal2011_HMC}. When $\epsilon$ is set too large, the discretization of the leapfrog integrator leads to inaccuracy which results in a high rejection rate for proposals. On the other hand, if $\epsilon$ is too small, the leapfrog integrator will take many small steps, leading to extended simulation times. The objective, therefore, is to find a balance that maintains an appropriate rate of acceptance. Problems can also arise from the choice of $L$. If $L$ is set too large, the algorithm may become computationally expensive and could miss important features of the target distribution because it overshoots regions of high probability. Conversely, if $L$ is too small the algorithm can exhibit a random walk behaviour and the generated samples may suffer from high autocorrelation.
 
Tuning these parameters can be quite difficult especially for complex models. Therefore, \textcite{Hoffman2014nuts} introduced the \textit{No-U-Turn sampler} (NUTS), which is an extension to the HMC algorithm that automatically optimizes $\epsilon$ and $L$.  
Their algorithm achieves this goal by adapting $\epsilon$ toward a target acceptance statistic during the initial warm-up phase. This approach allows for a broader exploration of the parameter space at the start, followed by a more refined search as convergence is reached. NUTS employs dual averaging to identify a suitable shrinkage factor that achieves a specified target acceptance rate.
Dual averaging is a prevalent method for tuning hyperparameters in algorithms that simultaneously considers both primal and dual spaces during the optimization process (see \textcite{nesterov2009primal} for a comprehensive explanation).

\textcite{Hoffman2014nuts} implement the No-U-Turn criterion to adaptively chose $L$ by simulating the Hamiltonian dynamics until a U-turn is detected. This means that the algorithm continues until it detects that the trajectory is about to double back on itself. To accomplish this, they build a binary tree of leapfrog steps, extending the trajectory in both forward and backward directions in time until a U-turn is detected. The depth of this tree determines the effective value of $L$.

 
\subsection{Variational Inference}
 
Like MCMC, VI aims to approximate the posterior distribution $p(\theta|X)$, where $\theta = \{\theta_1, \dots, \theta_K\}$ represents the latent variables of interest and $X = \{x_1, \dots, x_N\}$ represents the observed data. However, unlike the former, VI fits a function of the variational family as closely as possible to the posterior of interest, rather than sampling from the exact posterior distribution \parencite{Blei_et_al_2017}. 

The idea of VI is to pick a family of distributions $q(\theta | \nu)$ (henceforth $q(\theta)$) over these latent variables with its own variational parameters $\nu$ and then find a setting of $\nu$ that approximates $p(\theta|X)$ in the best way possible. Therefore, VI transforms inference to an optimization problem. It is crucial for this approach to be able to determine the distance between two distributions. 

The \textit{Kullback-Leibler} (KL) divergence is an information-theoretical measure that can be used to quantify that distance \parencite{Jordan1999introduction}.  In VI the continuous KL-divergence is defined as
\begin{align} \label{eq: KL}
	D_{KL}\big(q(\theta) \ || \ p(\theta|X)\big)= \int q(\theta) \ log \ \frac{q(\theta)}{p(\theta|X)} \ d\theta,
\end{align}
which is also known as \textit{reverse} KL-divergence or \textit{information projection} \parencite{Murphy2012machine}. It is non-negative and  in contrast to other distance measures not symmetric, so in general 
\begin{align}
	D_{KL}\big(q(\theta) \ || \ p(\theta|X)\big) \neq D_{KL}\big(p(\theta|X) \ || \ q(\theta)\big)
\end{align} 
holds \parencite{Cover_Thomas_2006}.\footnote{The \textit{forward} KL-divergence $D_{KL}\big(p(\theta|X) \ || \ q(\theta)\big)$ is used in another approximation technique known as \textit{expectation propagation}, where it is minimized locally between tilted distributions and their approximations \parencite{Minka2001_EP}.} 
By applying expectations with respect to $q(\theta)$ (denoted as $\mathbb{E}_{q(\theta)}$), equation \eqref{eq: KL} can be rewritten as follows:
\begin{align*}
	D_{KL}\big(q(\theta) \ || \ p(\theta|X)\big) &= \mathbb{E}_{q(\theta)}\Bigg[log  \ \frac{q(\theta)}{p(\theta|X)}\Bigg] \\
	&= \mathbb{E}_{q(\theta)}\big[log \ q(\theta)\big] - \mathbb{E}_{q(\theta)}\big[log \ p(\theta,X)\big] + \mathbb{E}_{q(\theta)}\big[log \ p(X)\big] \\
	&= \mathbb{E}_{q(\theta)}\big[log \ q(\theta)\big] - \mathbb{E}_{q(\theta)}\big[log \ p(\theta,X)\big] + log \ p(X). \numberthis \label{eq: KL_expect}
\end{align*}

As stated before, the goal is to find the member of the family that minimizes the KL-divergence to the exact posterior
\begin{align}
	q^*(\theta) = \underset{q(\theta)}{\text{arg min}} \  D_{KL}\big(q(\theta) \ || \ p(\theta|X)\big).
\end{align}

But directly evaluating equation \eqref{eq: KL_expect} requires the calculation of the logarithm of the evidence $ log \ p(X)$ that is generally intractable \parencite{Blei_et_al_2017}. However, since this term is constant with respect to $q(\theta)$, it can be omitted for optimization. Since the logarithm is a concave function, Jensen's inequality can be applied, which asserts that for any concave function, the expectation of the function is always less than or equal to the function's value at the expectation \parencite{Ganguly2021introduction}. This inequality allows to derive the so-called \textit{evidence lower bound} (ELBO), a key concept in VI that provides a tractable lower bound for the otherwise intractable log evidence:

\begin{align*}
	log \ p(X) &= log \int p(X, \theta) \ d\theta \\
	&= log  \int p(X, \theta) \frac{q(\theta)}{q(\theta)} \ d\theta \\
	&= log \ \mathbb{E}_{q(\theta)}\Bigg[\frac{p(X, \theta)}{q(\theta)}\Bigg] \\
	&\ge \underbrace{\mathbb{E}_{q(\theta)}\big[log \ p(X, \theta)\big] - \mathbb{E}_{q(\theta)}\big[log \ q(\theta)\big]}_{= ELBO(q(\theta))}. \numberthis  \label{eq: ELBO}
\end{align*}
Comparing the ELBO with equation \eqref{eq: KL_expect}, one can see that maximizing the ELBO is equivalent to minimizing the KL-divergence
\begin{align}
	D_{KL}\big(q(\theta) \ || \ p(\theta|X)\big) = -ELBO(q(\theta)) + log \ p(X).
\end{align}

The complexity of maximising the ELBO highly depends on the chosen variational family \parencite{Blei_et_al_2017}. One of the most widely used families is known as the \textit{mean-field} variational family, where $\theta$ is partitioned in its individual variables $\theta_k$ for $k = 1, \dots, K$. These components are assumed to be independent of each other. Therefore, each of the latent variables is assigned its own variational factor
\begin{align} \label{eq: var_fam}
	q(\theta) = \prod_{k=1}^{K} q_k(\theta_k).
\end{align}
Note that there is no specific parametric form for these factors and that the equation above does not depend on the observations $X$. More complex families include, for instance, structured or mixture-based variational families. The former allow for dependencies between the latent variables (\cite{SaulJordan1996structure} or \cite{BarberWiegerinck1999structure}) and the latter incorporate additional latent variables in the variational family \parencite{Bishop_et_al_1998mixture}. In this work, the focus will be on the mean-field family, chosen for its simplicity.

\subsubsection{Coordinate Ascent Variational Inference}
One commonly used algorithm to optimize the ELBO is the CAVI algorithm (\cite{Bishop2006}, \cite{Blei_et_al_2017}, \cite{Plummer_et_al_2020_CAVI}). The main idea behind CAVI is to iteratively update the parameters of the approximate variational distribution  by optimizing one variational factor $q_j = q_j(\theta_j)$ with $j \in K$, while holding the remaining factors $q_{i\neq j} $ fix for $i = 1, \dots, K$. Therefore, with the definition of the ELBO in equation \eqref{eq: ELBO} and the mean-filed assumption in equation \eqref{eq: var_fam}, the objective function becomes:
\begin{align} \label{eq: ELBO_qj}
	ELBO(q_j) = \int \big[q_j \prod_{i \neq j}^{K} q_i\big] \big[log \ p(X, \theta) - (log \ q_j + \sum_{i\neq j}^{K} log \ q_i )\big] d\theta.
\end{align} 

It can be demonstrated that the equation presented above attains a proportional maximization through the application of the exponentiated expected logarithm of the joint distribution\footnote{A comprehensive derivation of this result can be found in Appendix \ref{sec: App_A}.}
\begin{align} \label{eq: qj_opt}
	q_j^*(\theta_j) = \frac{exp\big(\mathbb{E}_{q_{i\neq j}}\big[log \ p(X, \theta_{i\neq j}, \theta_j)\big]\big)}{\int exp\big(\mathbb{E}_{q_{i\neq j}}\big[log \ p(X, \theta_{i\neq j}, \theta_j)\big]\big) d\theta_j}  \propto exp\big(\mathbb{E}_{q_{i\neq j}}\big[log \ p(X, \theta_{i\neq j}, \theta_j)\big]\big).
\end{align}

To iteratively update the variational distribution, the exponentiated expected logarithm of the full conditional distribution is used, which is also proportional to \eqref{eq: qj_opt}:
\begin{align*}
		q_j^*(\theta_j) &  \propto exp\big(\mathbb{E}_{q_{i\neq j}}\big[log \ p(X, \theta_{i\neq j}, \theta_j)\big]\big)\\
		&= exp\big(\mathbb{E}_{q_{i\neq j}}\big[log \ \big(p(\theta_j |X, \theta_{i\neq j}) \ p(X, \theta_{i \neq j})\big)\big]\big) \\
		&= exp\big(\mathbb{E}_{q_{i\neq j}}\big[log \ p(\theta_j |X, \theta_{i\neq j})\big] + \mathbb{E}_{q_{i\neq j}}\big[log \ p(X, \theta_{i \neq j})\big]\big) \\
		&= exp\big(\mathbb{E}_{q_{i\neq j}}\big[log \ p(\theta_j |X, \theta_{i\neq j})\big]\big) exp\big( \mathbb{E}_{q_{i\neq j}}\big[log \ p(X, \theta_{i \neq j})\big]\big) \\
		& \propto exp\big(\mathbb{E}_{q_{i\neq j}}\big[log \ p(\theta_j |X, \theta_{i\neq j})\big]\big). \numberthis \label{eq: qj_cond}
\end{align*} 

In the application of CAVI, convergence is typically monitored by observing changes in the ELBO \parencite{Blei_et_al_2017}. Convergence is declared once the fluctuations in the ELBO fall beneath a predetermined, minimal threshold $\delta$. 
In summary, the algorithm can be broken down into the following steps:
\begin{enumerate}
	\item Initialize variational factors $q_j(\theta_j)$.
	\item While the change in the ELBO is greater than $\delta$: 
	
	Iterate through all $j \in \{1, \dots, K\}$ and calculate
	$$ q_j(\theta_j) \propto exp\big(\mathbb{E}_{q_{i\neq j}}\big[log \ p(\theta_j |X, \theta_{i\neq j})\big]\big).$$
	
	 Compute the ELBO according to equation \eqref{eq: ELBO_qj}.
	\item Return variational density $q(\theta) = \prod_{j=1}^K q_j(\theta_j)$.
	
\end{enumerate}

The coordinate ascent structure of the algorithm requires iterating through the entire dataset at each iteration,  a process that can be computationally demanding. This makes the algorithm less appealing for large-scale applications. Because CAVI requires the computation of the conditional posterior distribution, it is especially suited for conditionally conjugate models, where a closed form solution exists \parencite{Dey2022robust}. A conditionally conjugate model refers to a statistical model where the prior distribution $p(\theta)$ of a parameter is conjugate to the conditional likelihood of that parameter given the other parameters or latent variables in the model \parencite{Gelman_et_al_2014}. This conditional conjugacy ensures that the conditional posterior distribution is in the same family as the prior distribution.

\subsubsection{Automatic Differentation Variational Inference} \label{sec: ADVI} 

\textcite{kucukelbir2017automatic} developed ADVI with the goal to establish a user-friendly method for creating variational inference algorithms that can be applied to a wide range of complex probabilistic models, including those that are not conditionally conjugate. The key idea in their approach is transforming all parameters, denoted as $\theta$, whether they are constrained or unconstrained, into a shared unconstrained real-valued parameter space. These transformed parameters are referred to as $\zeta = T(\theta)$, where $T: supp(p(\theta)) \rightarrow \mathbb{R}^K$ is a injective differentiable function. This function maps the support of the original parameter space into a $K$-dimensional real coordinate space. As a result, the optimization problem with respect to the ELBO is performed within this real coordinate space. In this context, the ELBO expression given in equation \eqref{eq: ELBO} needs to be redefined within the domain of the real coordinate space:
\begin{align}
	ELBO\big(q(\zeta)\big) = \mathbb{E}_{q(\zeta)}\big[log \ p(X, \zeta)\big] - \mathbb{E}_{q(\zeta)}\big[log \ q(\zeta)\big].
\end{align}

The joint density $p(X, \zeta)$ is the product of the joint density in the original parameter space with $\theta = T^{-1}(\zeta)$ and the absolute value of the Jacobian of the inverse of $T$:
\begin{align*} \numberthis \label{eq: joint_dens}
	p(X, \zeta) &= p(X, T^{-1}(\zeta)) \big|  \det \ J_{T^{-1}} (\zeta)  \big|, \\ \\
	&\text{with } J_{T^{-1}(\zeta)} = \begin{pmatrix}
		\frac{\partial T_1^{-1}}{\partial \zeta_1} & \cdots & \frac{\partial T_1^{-1}}{\partial \zeta_K} \\
		\vdots & \ddots & \vdots \\
		\frac{\partial T_K^{-1}}{\partial \zeta_1}  &\cdots & \frac{\partial T_K^{-1}}{\partial \zeta_K} 
	\end{pmatrix}. 
\end{align*}

The Jacobian adjustment is needed to guarantee that the density integrates to one after the transformation. Through this transformation the application of a single variational family across different models is possible, thereby enhancing the algorithm's universality.
As variational family \textcite{kucukelbir2017automatic} apply a factorized Gaussian family
$
	q(\zeta| \phi) = \prod_{k=1}^{K} \mathcal{N}(\zeta_k | \mu_k, \sigma_k^2),
$
where $\phi = (\mu_1, \dots, \mu_K, \sigma_1^2, \dots, \sigma_K^2)$ are the variational parameters of the variational distribution. Note that here $q(\zeta|\phi)$ does not denote the conditional distribution, but describes that the distribution $q(\zeta)$ is parametrized by $\phi$.  Because the variance parameters must always be positive, the standard deviations must be transformed to $\omega_k= log(\sigma_k)$ for $k = 1, \dots, K$. This transformation ensures that the support  of $\omega$ is on the real coordinate space and $\sigma$ is always positive. The factorized Gaussian variational approximation becomes
\begin{align}\label{eq: omega_Gauss}
	q(\zeta| \phi) = \prod_{k=1}^{K} \mathcal{N}(\zeta_k | \mu_k, exp(\omega_k)^2),
\end{align}
with $\phi = (\mu_1, \dots, \mu_K, \omega_1, \dots, \omega_K)$.
It is noteworthy that ADVI can be used with a full-rank Gaussian $p(\zeta|\phi) = \mathcal{N}(\zeta|\mu, \Sigma)$ variational family, where $\Sigma$ represents the complete covariance matrix. Although this approach can provide a more flexible approximation, it leads to a substantial increase in computational costs. Therefore, this work will focus on the mean-field Gaussian from equation \eqref{eq: omega_Gauss}.
With equation \eqref{eq: joint_dens} the ELBO in the real coordinate space becomes

\begin{multline} \label{eq: ELBO_real_space}
	ELBO\big(q(\zeta|\phi)\big) = \mathbb{E}_{q(\zeta|\phi)}\Big[log \ p(X, T^{-1}(\zeta)) + log \ \big|  \det \ J_{T^{-1}} (\zeta)  \big|\Big]  \\ 
	\underbrace{- \mathbb{E}_{q(\zeta|\phi)}\big[log \ q(\zeta|\phi)\big]}_{\mathbb{H}(q(\zeta| \phi))},
\end{multline}
where $\mathbb{H}(q(\zeta| \phi))$ is the entropy defined by \textcite{Shannon1948mathematical}. Following \textcite{Chen2016entropy} the entropy of a multivariate Gaussian in $\mathbb{R}^K$ is given by
\begin{align}
	\mathbb{H}(q(\zeta|\phi)) = \frac{K}{2} log(2\pi) + \frac{K}{2} + \frac{1}{2} \  log\big(\text{det} \ \Sigma \big).
\end{align}

Because $q(\zeta|\phi)$ is a factorized Gaussian, the determinant of the covariance matrix is $\text{det} \ \Sigma = \text{det} \ diag(exp(\omega)^2) = \prod_{k=1}^K \exp(\omega_k)^2$ and the entropy becomes
\begin{align*}
	\mathbb{H}(q(\zeta|\phi)) &= \frac{K}{2}\big(1 + log(2\pi)\big) + \frac{1}{2} \  log\Big(\prod_{k=1}^K exp(\omega_k)^2 \Big)\\
	&= \frac{K}{2}\big(1 + log(2\pi)\big) + \frac{1}{2} \  \sum_{k=1}^K log\big(exp(\omega_k)^2 \big) \\
	&= \frac{K}{2}\big(1 + log(2\pi)\big) + \sum_{k=1}^K \omega_k. \numberthis\label{eq: H}
\end{align*}

\begin{figure}[H] 
	\begin{center}
		\vspace{-2ex}
		\includegraphics[width=1\textwidth]{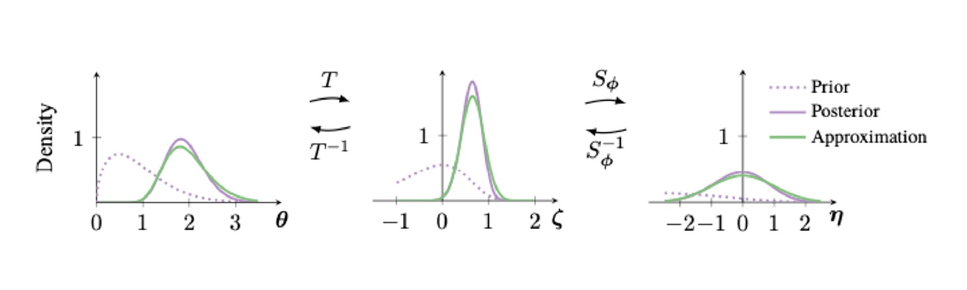}  
	\end{center}
	
	\caption{Illustration of the transformation process from the (constrained) parameter space to the real coordinate space and to the standardized space as applied in ADVI. Depicted as in \textcite{kucukelbir2017automatic}. }
	\label{abb: advi}	
\end{figure}

Unfortunately, the expectation in  equation \eqref{eq: ELBO_real_space} remains intractable. \textcite{kucukelbir2017automatic} address this issue by proposing an additional transformation they termed as \textit{elliptical standardization} to the variational family. This transformation is recognized in the literature as the \textit{Mahalanobis transformation} \parencite{Haerdle2012}, the \textit{re-parameterization trick} \parencite{Kingma2022autoencoding} or the \textit{coordinate transformation} \parencite{Rezende2015variational} and ensures that the expectation can be efficiently estimated using Monte Carlo integration.
The standardization $\eta = S_\phi(\zeta) = \text{diag}(exp(\omega))^{-1} (\zeta - \mu) $ absorbs the variational parameters $\phi$, leading to a standard Gaussian variational distribution
\begin{align}
	q(\eta) = \mathcal{N}(\eta| 0, I) = \prod_{k=1}^K \mathcal{N}(\eta_k|0, 1), 
\end{align}
where $I$ is the identity matrix. The entire transformation process of ADVI from the (constrained) parameter space through the real coordinate space to the standardized space is depicted in figure \ref{abb: advi}. The approximation using the variational density is represented by the green line. 

With these transformations and incorporating $\omega= log(\sigma)$ into equation \eqref{eq: H}, the ELBO in the standardized space finally becomes
\begin{multline}
	ELBO(q(\eta)) = \mathbb{E}_{q(\eta)}\Big[log \ p\Big(X, T^{-1}\big(S_{\phi}^{-1}(\eta)\big)\Big) + log \ \big|det \ J_{T^{-1}}\big(S_{\phi}^{-1}(\eta)\big) \big|\Big] \\ + \frac{K}{2}\big(1 + log(2\pi)\big) + \sum_{k=1}^K \omega_k. 
\end{multline}

ADVI solves the optimization problem 
\begin{align}
	\mu^*, \omega^* = \underset{\mu, \omega}{\text{arg max}} \ ELBO(q(\eta))
\end{align} 
with stochastic gradient ascent and an adaptive step-size sequence that ensures convergence . Therefore, the calculation of the gradients of the ELBO is required. Because gradients can be described as limits, with the dominated convergence theorem, one can interchange the order of taking gradients and expectations \parencite{Cinlar2011}. Applying this result and the chain rule leads to the following gradients of the ELBO:
\begin{align*}\label{eq: grad_mu}
	\nabla_\mu ELBO(q(\eta)) &= \frac{\partial ELBO(q(\eta))}{\partial \mu} \\
	&= \mathbb{E}_{q(\eta)}\Big[\nabla_\theta log \ p(X, \theta) \nabla_\zeta T^{-1}(\zeta) + \nabla_\zeta log \ | det \ J_{T^{-1}} (\zeta)|\Big], \numberthis 
\end{align*}
\begin{multline}\label{eq: grad_omega}
	\nabla_\omega ELBO(q(\eta))=  \mathbb{E}_{q(\eta)}\Big[\big(\nabla_\theta log \ p(X, \theta) \nabla_\zeta T^{-1}(\zeta) + \nabla_\zeta log \ | det \ J_{T^{-1}} (\zeta)|\big) \\ \cdot \eta^T diag(exp(\omega))\Big] +1.
\end{multline}

The gradients inside the expectations can be calculated using \textit{automatic differentiation} (AD) (see \textcite{Baydin2018AD} for a detailed review of AD).  The foundation of AD is the chain rule. AD breaks down complex functions into elementary operations (like addition or multiplication) and functions (like sine or cosine). It creates a computational graph where each node represents an elementary operation or function. Then it systematically applies the chain rule to compute derivatives. \textcite{kucukelbir2017automatic} use \textit{reverse mode} AD, where first, a forward trace through the computational graph from input to output is computed and the value of each intermediate node is stored. Afterwards, the derivatives of each node with respect to its inputs are computed in a reverse trace starting from the output. Reverse mode AD is especially efficient when the function has many input variables but only a few output variables \parencite{Baydin2018AD}. 

In conclusion, ADVI can be summarized in the following steps. While the ELBO has not converged in the $i$-th iteration, do:
\begin{enumerate}
	\item Draw $M$ samples $\eta_m \sim \mathcal{N}(0, I)$.
	\item Use Monte Carlo integration and AD to approximate $\nabla_\mu ELBO$ in equation \eqref{eq: grad_mu} and $\nabla_\omega ELBO$ in equation \eqref{eq: grad_omega}.
	\item Adaptively set stepsize $\rho^{(i)} $.
	\item Update $\mu^{(i + 1)} = \mu^{(i)} + \text{diag}(\rho^{(i)}) \nabla_\mu ELBO$.
	\item Update $\omega^{(i+1)} = \omega^{(i)} + \text{diag}(\rho^{(i)})\nabla_\omega ELBO$.
\end{enumerate}

	


\section{Models} \label{sec: models}

Many discrete choice models derive from the assumption that individuals want to maximize their utility, which represents the preference an individual has for each alternative. In addition, utility is subject to random variation, reflecting the presence of unobservable factors that influence decision making. Therefore, theses models are called \textit{random utility maximization} (RUM) models. They were introduced in the economic context by \textcite{Marschak1960rum}. An individual $i$ will choose product $j_c$ ($y_{ij_c} =  1$ and $0$ otherwise) from a given category $c$ if this product provides a higher utility $U_{ij_c}$than the other alternatives $l_c$:
\begin{align}
	y_{ij_c} = 1 \ \ if \ \ U_{ij_c} > U_{il_c} \ \ \forall j_c \neq l_c.
\end{align}

The individual specific utility of each product consists of a deterministic part $u_{ij_c} $ and an error term $\epsilon_{ij_c}$ capturing all factors that are not included in $u_{ij_c} $ but affect the utility
\begin{align}
	U_{ij_c} = u_{ij_c} + \epsilon_{ij_c}.
\end{align}

Under the assumption that $\epsilon_{ij_c}$ is\textit{ independent and identically distributed} (i.i.d.) of extreme value type 1, \textcite{McFadden1973logit} showed that the probability of individual $i$ choosing product $j_c$ is
\begin{align} \label{eq: choice_prob_rum}
	P_{ij_c} = p(y_{ij_c} = 1) = \frac{exp(u_{ij_c})}{\sum_{l_c=1}^{J_c} exp(u_{il_c})},
\end{align}
the so called \textit{conditional logit} probability. Note that the assumption of an i.i.d. extreme value type 1 distributed error term $\epsilon_{ij_c}$ will remain for the rest of this work.
If the deterministic part is specified as a linear model $u_{ij_c} = \beta' x_{ij_c}$ of parameter vector $\beta$ and observed variables $x_{ij_c}$, 
the model is known as the multinomial logit model \parencite{Train2003}. This model suffers from many limitations, especially the \textit{independence of irrelevant alternatives} (IIA) assumption (see \textcite{Train2003} for a detailed discussion). Under IIA the relative odds of selecting alternative $i$ over alternative $k$ remain constant, independent of the available alternatives or the attributes associated with those alternatives. For example, when considering demand in a supermarket, the assumption implies that if the price of a product j increases, customers would adjust their demand proportionally to their initial preferences. Intuitively, however, one would assume that demand for products similar to j would increase over-proportionally.

\subsection{Mixed Logit} \label{sec: MixedLogit}
The mixed logit model is a commonly used model for discrete choice estimation, which overcomes most of the limitations of the multinomial logit model \parencite{HensherGreene2003mixedLogit}. Because of its highly flexible structure, it allows incorporating random variations in preferences, unrestricted substitution patterns between choices, and accounting for correlations in unobservable factors over time \parencite{Train2003}. 
Therefore, it is able to approximate choice probabilities derived from any RUM model \parencite{McFaddenTrain2000mmnl}.

The deterministic part of the utility individual $i$ derives from choosing product $j_c$ is 
\begin{align}
	u_{ij_c} = \beta_i' \  x_{ij_c},
\end{align}
where $\beta_i'$ is a vector of individual specific random coefficients and $x_{ij_c}$ includes all observable variables for individual $i$ and product $j_c$. The density of this coefficients is called the mixing distribution and is denoted as $f(\beta | \theta)$, where $\theta$ are the parameters of the distribution \parencite{Train2003}. Conditional on $\beta_i$ the choice probability is equal to equation \eqref{eq: choice_prob_rum}. However, $\beta_i$ is not observable and therefore, one can not condition on it. Hence, the unconditional choice probability is obtained by integrating over all possible values of $\beta_i$:
\begin{align} \label{eq: choice_prob_mix}
	P_{ij_c} = \int_\beta \Bigg(\frac{exp(u_{ij_c})}{\sum_{l_c = 1}^{J_c} exp(u_{il_c})}\Bigg) \ f(\beta | \theta ) \ d\beta.
\end{align}
In many applications \parencite{BraunMcAuliffe2010}, the mixing distribution is assumed to be normal distributed $f(\beta | \theta) \sim \mathcal{N}(\mu, \sigma)$. Since $\beta$ is integrated out in the unconditional choice probability above, mean $\mu$ and standard deviation $\sigma$ are the parameters to be estimated. 

The integral in equation \eqref{eq: choice_prob_mix} cannot be solved analytically. Therefore, simulated \textit{maximum likelihood} (ML) is used to estimate the parameters. By taking $R$ draws from the mixing distribution with label $\beta^r$ for $r = 1, \dots, R$, the simulated choice probability becomes
\begin{align} \label{eq: sim_choice_prob}
	\widetilde{P}_{ij_c} =  \frac{1}{R}\sum_{r =1}^{R} \Bigg(\frac{exp(\beta^{r'}_i \ x_{ij_c})}{\sum_{l_c = 1}^{J_c} exp(\beta^{r'}_i \ x_{il_c})}\Bigg).
\end{align}
The \textit{simulated log likelihood} (SLL) is then given by
\begin{align} \label{eq: sim_logLik}
	SLL = \sum_{i = 1}^{I} \sum_{j_c = 1 }^{J_c} \mathbf{1}_{\{y_{ij_c} = 1\}} \ log  (\widetilde{P}_{ij_c}),
\end{align}
where $\mathbf{1}_{\{y_{ij_c} = 1\}}$ is the indicator function which equals 1 if individual $i$ choses $j_c$ and zero otherwise. The specified likelihood treats individual choices separately by integrating over the random coefficients for each choice. The maximization of the SLL can be solved numerically with several algorithms like \textit{Berndt-Hall-Hall-Hausman} (BHHH) \parencite{berndt1974estimation}, \textit{Broyden-Fletcher-Goldfarb-Shanno} (BFGS) (\cite{broyden1970convergence}, \cite{fletcher1970new}, \cite{goldfarb1970family}, \cite{shanno1970conditioning}) or the \textit{limited memory BFGS} (L-BFGS) \parencite{liu1989limited}.

\subsection{Latent Factorization Model} \label{sec: latent fac model}

Latent factorization models are quite popular in the machine learning literature, because they reduce the number of parameters to be estimated and allow for a highly flexible model structure (see \textcite{koren2009matrix} or \textcite{Athey_2019MLreview}). This work considers a factorization model which can be thought of a simplified version of the travel time factorization model by \textcite{Athey2018ttfm} or the product-level model of the nested factorization model by \textcite{Donnelly2021counterfactual}. \textcite{Athey_2019MLreview} note that the application of factorization techniques can enhance traditional consumer choice models. Therefore, the deterministic part of the utility derived by an individual $i$ from selecting product $j$ in category $c$ is expressed as
\begin{align*}
	u_{ij_c} &= \beta_i' \theta_{j_c}  - \gamma_i' \lambda_{j_c} \ p_{j_c} \numberthis \label{eq: fac_mod}\\
	&= \begin{pmatrix}
		\beta_{i1} &  \cdots & \beta_{iK}
	\end{pmatrix} \cdot \begin{pmatrix}
	\theta_{j_c1}\\
	\vdots \\
	\theta_{j_cK}
\end{pmatrix} - \begin{pmatrix}
\gamma_{i1} &  \cdots & \gamma_{iK}
\end{pmatrix} \cdot \begin{pmatrix}
\lambda_{j_c1}\\
\vdots \\
\lambda_{j_cK}
\end{pmatrix} \cdot p_{j_c},
\end{align*}
where $\beta_i$, $\theta_{j_c}$, $\gamma_i$ and $\lambda_{j_c}$ are $K$-dimensional vectors of latent variables. The vector $\theta_{j_c}$ can be interpreted as a vector of unobserved product characteristics for product $j_c$. Concurrently, $\beta_i$ represents the unobserved preferences of individual $i$ corresponding to these characteristics. $\gamma_i$ and $\lambda_{j_c}$ signify latent individual and product specific sensitivities for price $p_{j_c}$.
With this structure the model allows for rich individual and product specific heterogeneity. It assumes that the categories are non overlapping and only one product is bought per category, which is called the unit demand assumption. Moreover, the model draws inferences from products that have similar choice patterns across individuals, as well as individuals who have similar choice patterns across products. This structure makes it suitable for a potential efficiency gain when multiple categories are considered at the same time. The choice probability conditional on the decision to buy one product from a given category follows the logit model \eqref{eq: choice_prob_rum}
\begin{align}
	p\Big(y_{ij_c = 1} \Big| \sum_{j = 1}^{J_c} y_{ij} = 1\Big) =  \frac{exp(u_{ij_c})}{\sum_{j_c=1}^{J_c} exp(u_{ij_c})}.  \label{eq: factor_prob}
\end{align}

Estimation of a latent factorization model can be computationally demanding because of the potentially large number of latent variables. Therefore, Bayesian machine learning methods like MCMC and VI are used for estimation in most applications (\textcite{Athey2018ttfm}, \textcite{Ruiz2020shopper} or \textcite{Donnelly2021counterfactual}).

\section{Simulation study} \label{sec: study}

In the current simulation study, discrete choice data is generated under various scenarios. The primary objective is to analyze and compare the computational efficiency and accuracy of MCMC and VI for an increasing amount of categories. Moreover, these techniques are benchmarked against the established mixed logit model to provide a holistic evaluation of their performance.

To limit the complexity of parameter estimation, only the price of a product is considered as a variable during the simulation. As a result, the deterministic part of the latent factorization model shrinks to 
\begin{align} \label{eq: fac_simulation}
	u_{ij_c} = \begin{pmatrix}
		\gamma_{i1} &  \cdots & \gamma_{iK}
	\end{pmatrix} \cdot \begin{pmatrix}
		\lambda_{j_c1}\\
		\vdots \\
		\lambda_{j_cK}
	\end{pmatrix} \cdot p_{j_c}
\end{align}
and the deterministic part of the mixed logit model becomes
\begin{align} \label{eq: mixedL_simulation}
	u_{ij_c} = \beta_i' \ p_{ij_c}.
\end{align}

\subsection{Data Generating Process and Settings}

This section provides details about the characteristics and chosen parameters of the DGP, based on which data sets are generated.
For this simulation study discrete choice data is generated according to a latent factorization model of dimension $K = 3$ for multiple categories. The utility individual $i$ gains from buying product $j_c$ from category $c$ is simulated according to equation \eqref{eq: fac_simulation} as

\begin{align}
	U_{ij_c} =  - \begin{pmatrix}
		\gamma_{i1} &  \gamma_{i2} & \gamma_{i3}
	\end{pmatrix} \cdot \begin{pmatrix}
		\lambda_{j_c1}\\
		\lambda_{j_c2}\\
		\lambda_{j_c3}
	\end{pmatrix} \cdot p_{j_c} + \epsilon_{ij_c}.
\end{align}

The negative sign indicates that as prices increase, the utility diminishes. Hence, only positive parameters are i.i.d. sampled from the following distributions:
\begin{align*}
	\gamma_{i1} &\sim exp\Big(\frac{1}{1.5}\Big)& & \lambda_{j_c1} \sim exp\Big(\frac{1}{2}\Big) \\
	\gamma_{i2} &\sim log \ \mathcal{N}(0, 1)& & \lambda_{j_c2} \sim  log \ \mathcal{N}(1, 0.5)  \numberthis \\
	\gamma_{i3} &\sim log \ \mathcal{N}(0, 0.5)& & \lambda_{j_c3} \sim  exp\Big(\frac{1}{0.65}\Big).
\end{align*}
This leads to substantial individual and product specific heterogeneity in the latent preferences. For instance, on the individual side, these latent factors might be interpreted as sensitivity to quality, brand loyalty or environmental consciousness. Correspondingly, on the product side, they can be perceived as the quality of the product, the strength of the brand  and its sustainability. Prices are uniformly drawn on the interval $[0, 5]$ and $\epsilon_{ij_d}$ is Gumbel distributed as stated before.

In the baseline setting, the DGP generates choice probabilities based on equation \eqref{eq: factor_prob} for a total amount of $I = 20$ individuals, each having the option to select one of $J_c = 5$ products within a category $c$. It's not mandatory for an individual to pick a product from each available category. Rather, the actual categories from which a product is chosen by an individual are randomly decided. To ensure an adequate number of observations, individuals must select from at least half of the total available categories $C_{min} = \lfloor \frac{C}{2} \rfloor$. To examine the possible efficiency improvements from observing multiple categories simultaneously, fifteen distinct category settings are modeled $C \in \{2, 3, 4, 5, 10, 20, 50, 100, 200, 300, 400, 500, 600, 700, 800\}$. To account for the randomness inherited in the DGP, 50 data sets are simulated for each of the 15 category settings. 

In the second setting, the number of individuals is increased to $I = 40$, while holding all other variables fix. As a result, a larger set of parameters need to be estimated, making the estimation process more computationally demanding.

For the third setting, the distributions of the parameters are adjusted to
\begin{align*}
	\gamma_{i1} &\sim exp\Big(\frac{1}{0.5}\Big)& & \lambda_{j_c1} \sim exp\Big(\frac{1}{0.5}\Big) \\
	\gamma_{i2} &\sim log \ \mathcal{N}(0, 0.5)& & \lambda_{j_c2} \sim  log \ \mathcal{N}(1, 0.25)  \numberthis \\
	\gamma_{i3} &\sim log \ \mathcal{N}(0, 0.25)& & \lambda_{j_c3} \sim  exp\Big(\frac{1}{0.55}\Big),
\end{align*}
which leads to lower individual and product specific heterogeneity in the latent preferences and a changed mean magnitude of price sensitivity.

\subsection{Computational Details}

The computations of MCMC and VI are executed using the probabilistic programming language Stan \parencite{carpenter2017stan}. In the case of MCMC, the latent factorization model is estimated with the adaptive path length setting HMC extension NUTS, as elaborated in section \ref{sec: HMC}. The algorithm is run with four Markov chains and 1000 samples are generated for each chain after the warm-up phase. The maximum depth of the binary tree used in NUTS is set to 10, resulting in a maximum of $2^{10}$ leapfrog steps per iteration.

For VI, the ADVI algorithm, detailed in section \ref{sec: ADVI}, is employed. To match the number of generated samples of MCMC, 4000 samples are drawn from the approximated variational density. As variational family a factorized Gaussian is applied as described in equation \eqref{eq: omega_Gauss}.

As previously noted, the mixed logit model is typically estimated via simulated ML. This estimation can be carried out in R with the \texttt{maxLik} package \parencite{maxLik2011}. However, this approach is computationally demanding. Given this challenge, this simulation study utilizes the optimization function provided by Stan to estimate the mixed logit model. When a Jacobian adjustment is implemented, the optimization function in Stan yields the mode of the posterior distribution \parencite{standev2023reference}. Without the Jacobian adjustment, the function returns the \textit{penalized maximum likelihood} (PML) estimate \parencite{carpenter2017stan}. The penalty term is the logarithm of the prior distribution \parencite{CousineauHelie2013}. As the sample size increases, the PML estimate converges towards the ML estimate. 
The optimization employs the L-BFGS algorithm, with a history memory size set to 5. In the computation of the simulated choice probabilities as outlined in equation \eqref{eq: sim_choice_prob}, $R = 1000$ i.i.d. draws are taken from a standard normal distribution.
Standard normal priors are set for all parameters. Computations are performed on the HILBERT cluster at the University of Düsseldorf.

\subsection{Results}

The factorization model is based on the dot product of $\gamma_i'\lambda_{j_c}$, which introduces several invariances, including label switching, rescaling and rotations of the latent factors \parencite{Donnelly2021counterfactual}. More generally, transformations of the latent factors that preserve their inner products leave the resulting utilities unchanged. Consequently, the individual latent parameters cannot be uniquely identified. Therefore, the emphasis shifts to assessing how accurately choice probabilities can be predicted rather than recovering the individual latent factors.

To account for the inherent randomness of the DGP, 50 datasets are simulated for each category setting. To evaluate the out-of-sample predictive performance of the algorithms and models, the data is partitioned into two separate subsets: a training dataset for parameter estimation and a test dataset for predictive assessment. In this simulation study, the training data consists of 80\% of the total generated observations, which are randomly selected. Consequently, the remaining 20\% of the total generated observations constitute the test dataset. Segmenting data in this manner is a common practice in machine learning literature, aiming to prevent overfitting and produce more general results. It is important to note that there is no need for a validation dataset in this simulation study, because hyperparameters are fixed independently of the test errors.

The prediction accuracy is measured for each generated dataset $d = 1, \dots, 50$ by the \textit{ root mean squared error} (RMSE)  across the number of observations $N$ in the test dataset and products per category $j_c = 1, \dots, J_c$:
\begin{align} \label{eq: RMSE_d}
	RMSE_d &= \sqrt{\sum_{n=1}^{N \cdot J_c} \frac{(\hat{p}_{n,j_c} - p^{\textit{true}}_{n,j_c})^2}{N \cdot J_c}},
\end{align}
were $\hat{p}_{n,j_c}$ represents the predicted choice probability that the individual in the $n$-th observation of the test set choses product $j_c$.
The reported RMSEs in the following figures are the average RMSEs over all dataset  and equal
\begin{align}
	RMSE_{avg} = \frac{1}{50} \sum_{d=1}^{50} RMSE_d
\end{align}  for the given settings. 

The applied Bayesian methods MCMC and VI return $S = 4000$ samples for each parameter of the posterior. Each of this samples $s = 1, \dots, S$ is used to make a prediction $\hat{p}_{s,n,j_c}$, which is the predicted choice probability that the individual in the $n$-th observation of the test set choses product $j_c$ based on the $s$-th sample of posterior parameters. The RMSE based on the $s$-th sample is
\begin{align}
	RMSE_s =  \sqrt{\sum_{n=1}^{N \cdot J_c} \frac{(\hat{p}_{s,n,j_c} - p^{\textit{true}}_{n,j_c})^2}{N \cdot J_c}}.
\end{align}

\begin{figure}[H] 
	\begin{center}
		
		\includegraphics[width=0.8\textwidth]{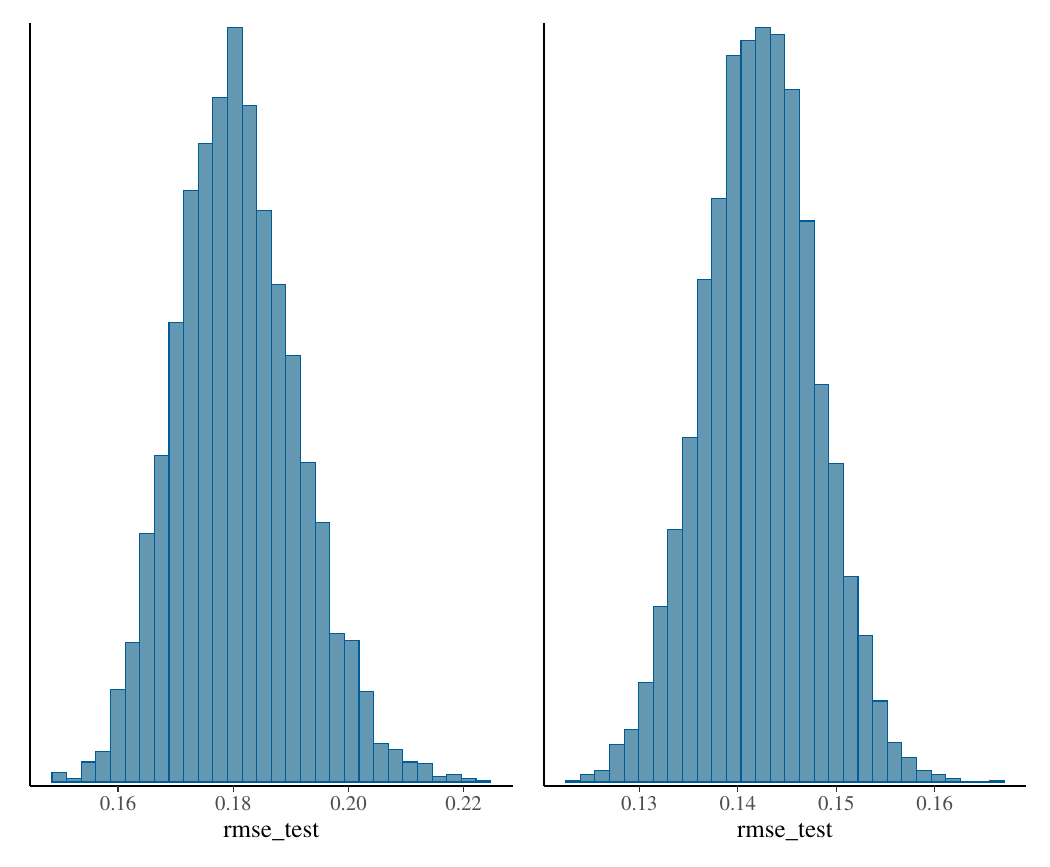}  
	\end{center}
	\vspace{-1ex}
	\caption{Example of $RMSE_s$ for $s=1,\dots, S$ estimated with VI (left panel) and MCMC (right panel) with $K = 3$ based on a DGP with $I=20$, $J_c=5$ and $C=200$.}
	\label{abb: rmse_hist_plt}	
\end{figure}

The variation in $RMSE_s$ across all posterior draws $S$ reflects uncertainty in predictive performance arising from parameter uncertainty conditional on the observed data set. However, it does not capture the sampling uncertainty of out-of-sample predictive performance. Figure \ref{abb: rmse_hist_plt} illustrates the uncertainty in the estimated $RMSE_s$ of VI (left panel) and MCMC (right panel) for an exemplary dataset with $I=20$ individuals, $C=200$ categories and $J_c = 5$ products per category.
The RMSE over the whole simulated dataset becomes $RMSE_d = \frac{1}{S}\sum_{s=1}^{S} RMSE_s$.

The PML estimate of the mixed logit model only returns point estimates of the parameters. Therefore, prediction accuracy is exactly measured as described in equation \eqref{eq: RMSE_d}.

\begin{figure}[H] 
	\begin{center}
		\vspace{-6ex}
		\includegraphics[width=0.9\textwidth]{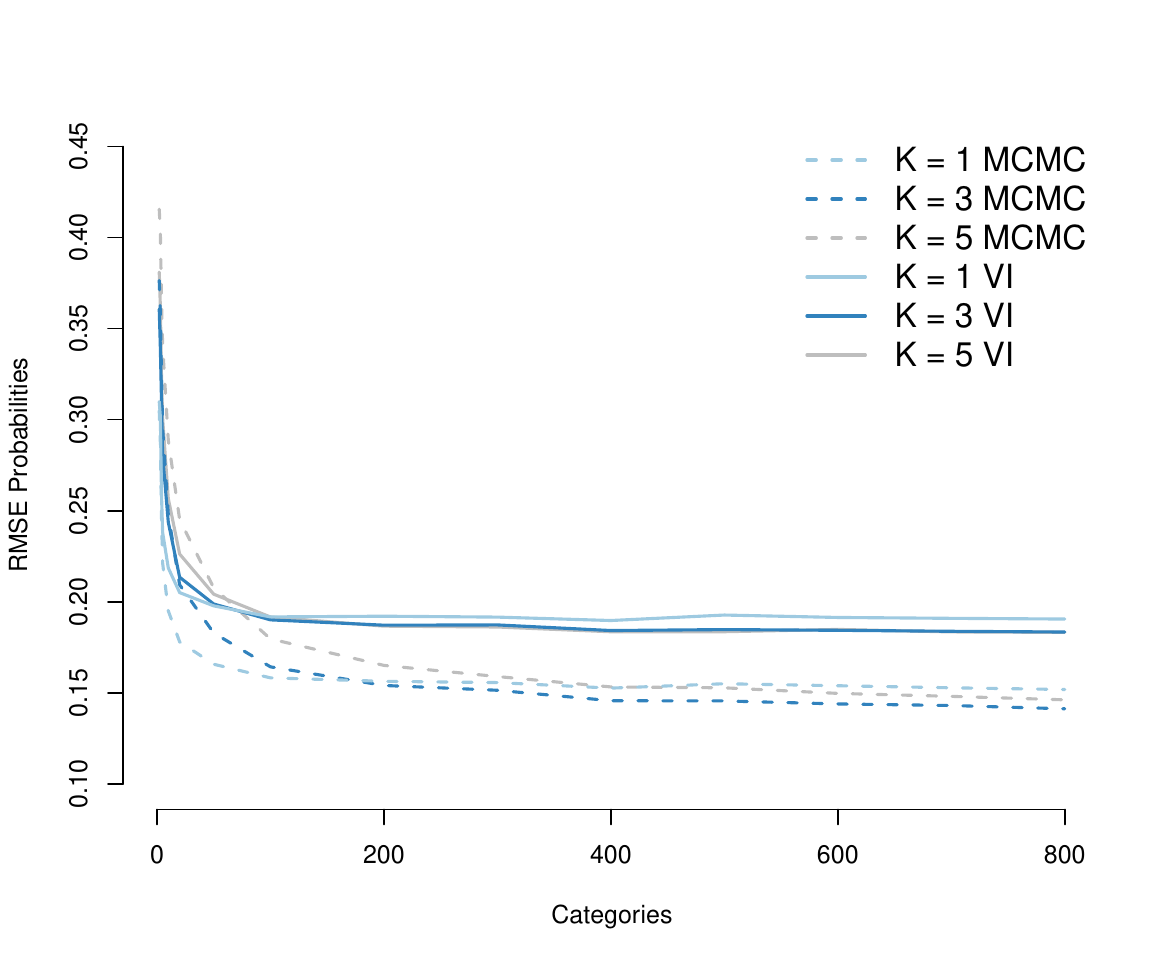}  
	\end{center}
	\vspace{-3ex}
	\caption{$\text{RMSE}_{avg}$ of the latent factorization model estimated with $K=1$, $K=3$ and $K=5$ factors for MCMC and VI.}
	\label{abb: results_k_plt}	
\end{figure}

While it is established for this simulation study that the data was generated using a latent factorization model with a dimension of $K=3$, figure \ref{abb: results_k_plt} illustrates the shifts in predictive accuracy when employing models with dimensions of $K=1$ or $K=5$. Notably, the model with $K=3$ yields the lowest RMSEs, especially as the number of categories increases, for both MCMC and VI. When analysing only a small set of categories, the one-dimensional model emerges as the most predictively accurate. Overall, MCMC's RMSEs are observed to be lower than those of VI. 

In terms of the runtimes of the different dimensional models, as observed in table \ref{tab:runtime_k}, both methods exhibit an increase in runtime as the dimension $K$ escalates. As anticipated, higher-dimensional models demand more computational resources, leading to longer runtimes. For the most complex model with $K=5$, the runtime of VI ranges from 5.27 seconds (for $C=2$) to 2.58 minutes (for $C=800$). In contrast, the runtime for MCMC spans from 0.34 minutes (for $C=2$) to roughly 3.25 hours (for $C=800$).

\begin{table}[H]
	\centering
	\caption{Average runtime in seconds for factorization model of different dimensions rounded to two decimal places.}
	\label{tab:runtime_k}
	\begin{tabular}{rrrrrrr}  
		\toprule
		&  \multicolumn{3}{c}{VI}& \multicolumn{3}{c}{MCMC}  \\
		C & $K =1$ & $K =3$ & $K =5$ & $K =1$ & $K =3$ & $K =5$ \\   
		\midrule
		 2 & 2.02 & 3.24 & 5.27 & 10.22 & 14.48 & 20.24 \\   
		3 & 2.35 & 3.79 & 5.39 & 10.99 & 17.13 & 22.13 \\   
		4& 2.50 & 4.11 & 5.75 & 12.62 & 20.60 & 27.31 \\   
		5 & 2.57 & 4.34 & 6.39 & 13.77 & 22.51 & 29.91 \\   
		10 & 3.62 & 6.14 & 8.34 & 22.55 & 38.12 & 53.94 \\   
		20 & 5.09 & 7.70 & 10.96 & 44.71 & 84.29 & 102.68 \\   
		50 & 9.72 & 14.90 & 21.53 & 139.09 & 222.51 & 337.50 \\   
		100& 15.57 & 23.11 & 39.52 & 361.58 & 712.71 & 1160.38 \\   
		200 & 24.46 & 39.78 & 62.08 & 862.10 & 1739.82 & 2667.04 \\   
		300 & 29.48 & 55.22 & 86.26 & 1851.44 & 2804.08 & 4306.88 \\   
		400 & 36.83 & 69.93 & 104.94 & 4069.11 & 4145.13 & 6421.57 \\   
		500& 37.34 & 78.53 & 116.93 & 4748.95 & 5031.88 & 8106.42 \\   
		600& 48.86 & 89.85 & 135.59 & 5130.75 & 6549.15 & 10193.17 \\   
		700 & 52.10 & 97.48 & 146.78 & 5004.50 & 7163.05 & 10849.70 \\   
		800 & 62.74 & 102.25 & 154.71 & 5417.82 & 8035.25 & 11712.74 \\    
		\bottomrule
	\end{tabular}
\end{table}

\begin{figure}[H] 
	\begin{center}
		
		\includegraphics[width=0.9\textwidth]{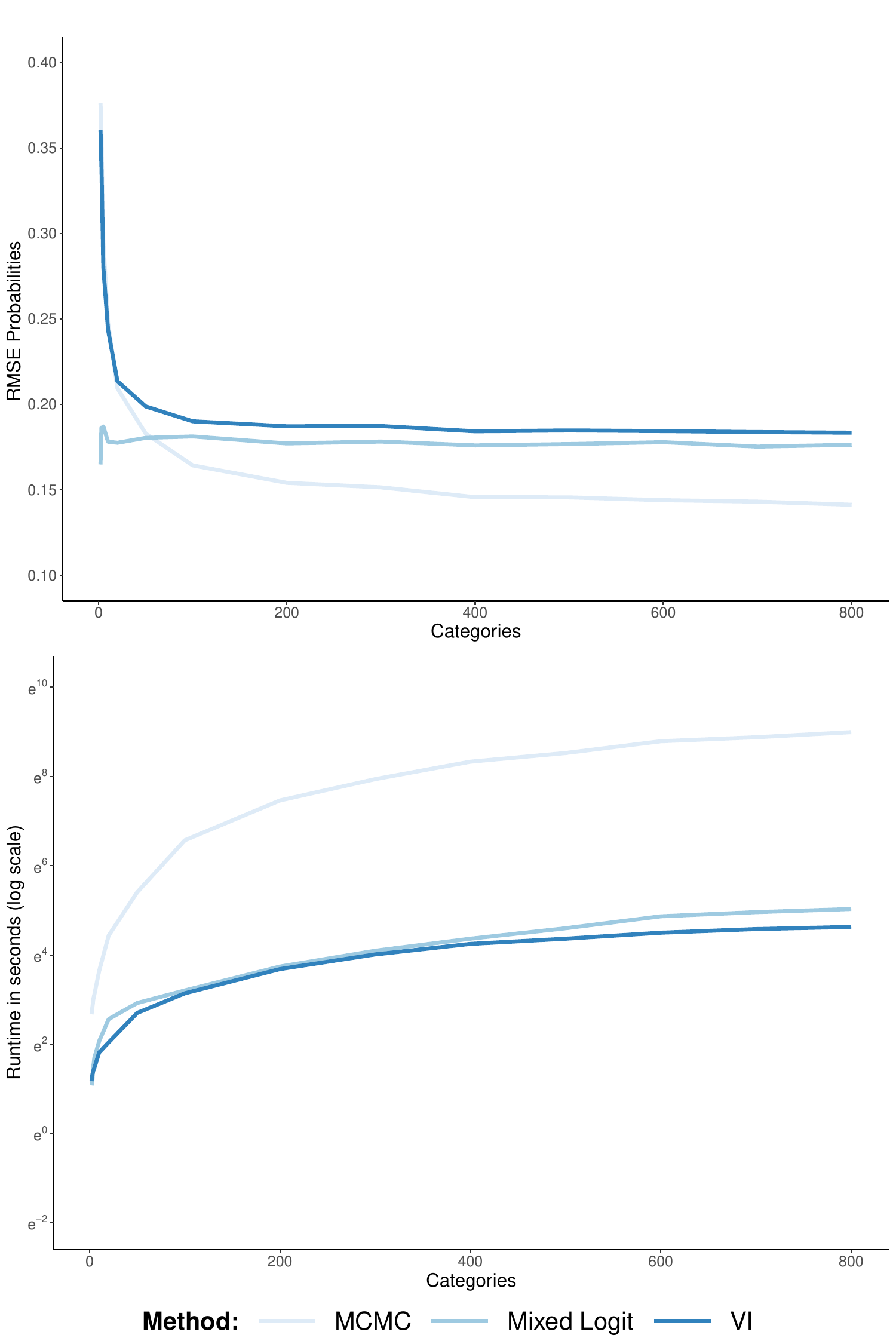}  
	\end{center}
	\vspace{-1ex}
	\caption{$\text{RMSE}_{avg}$ and runtime of the baseline setting with $K = 3$, $I=20$ and $J_c=5$.}
	\label{abb: results_k3_plt}	
\end{figure}

Figure \ref{abb: results_k3_plt} compares the latent factorization model, represented by MCMC and VI and the mixed logit model. The depicted results originate from the baseline setting featuring dimensions $K=3$, 20 individuals and 5 products within each category.
The upper panel shows the $\text{RMSE}_{avg}$ of the choice probabilities. MCMC and VI both have high average RMSEs if only two categories are considered with values of 0.376 and 0.381, respectively. However, a transformation in performance becomes evident as the amount of categories escalates.  For the maximum amount of considered categories, MCMC reaches a RMSE of 0.141 and VI a value of 0.183. 
In the beginning, with a small amount of categories, both methods perform equally badly. Yet, a shift occurs around the 10-category mark, where MCMC begins to overtake VI in terms of predictive accuracy. After observing more than 200 categories, both methods seem to converge to a RMSE of roughly 0.18 for VI and 0.15 for MCMC. In contrast to these two methods, the mixed logit model stays relative constant at a average RMSE of roughly 0.17, which indicated that the model does not gain efficiency with the amount of categories under the chosen DGP and model specification.

As depicted in the lower panel of Figure \ref{abb: results_k3_plt}, the runtimes of the methods are presented on a logarithmic scale to accommodate the disparity between the runtime of MCMC and the other approaches. Both the mixed logit and VI display nearly analogous runtimes. However, when considering an extensive number of categories, VI demonstrates a slight computational advantage, marking it as more efficient in these scenarios.
Most notably, the runtime of MCMC stands out. It increases substantially with the number of categories, exhibiting approximately superlinear growth over parts of the investigated range.. This trend is particularly pronounced in the category range spanning from 2 to 200.

\begin{figure}[H] 
	\begin{center}
		
		\includegraphics[width=1\textwidth]{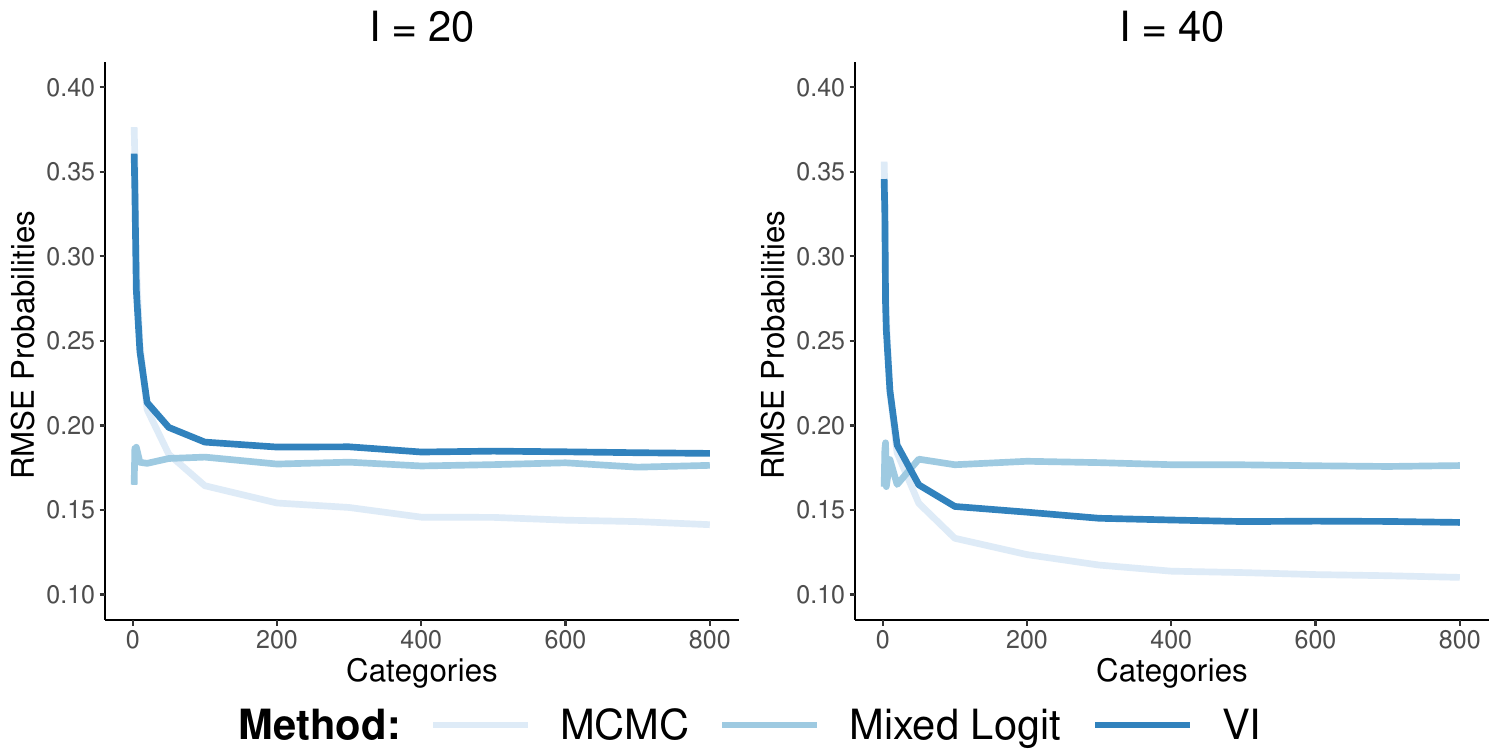}  
	\end{center}
	\vspace{-1ex}
	\caption{Comparison of $\text{RMSE}_{avg}$ for $I = 20$ and $I=40$. In both settings with $K=3$ and $J_c=5$.}
	\label{abb: results_i40_rmse_plt}	
\end{figure}

A comparison between the $\text{RMSE}_{avg}$ of the baseline setting and the second setting with an increased number of individuals  $I=40$ can be found in figure \ref{abb: results_i40_rmse_plt}. A prominent observation is the substantial reduction in RMSE values for the latent factorization model when the number of individuals is elevated to 40. MCMC reaches a lowest RMSE of 0.110 and VI achieves a lowest value of 0.143. In contrast, the predictive accuracy of the mixed logit model remains unchanged by the data augmentation. Its predictive accuracy appears to stagnate, with the RMSE consistently hovering around the 0.17 mark, mirroring its performance in the baseline scenario. 
The gained predictive accuracy comes with a trade-off in the runtime of the latent factorization model, as table \ref{tab:runtime_i40} shows. Doubling the number of individuals from 20 to 40 generally leads to an increase in runtime for all methods across all category settings. Similar to the baseline setting, MCMC is the most time-consuming method across all categories. For instance, at $C=2$,  MCMC takes 21.74 seconds, while VI and mixed logit complete in 4.87 and 5.06 seconds respectively. As the number of categories grows, the runtime disparity between MCMC and the other methods becomes even more pronounced. At $C=800$, the MCMC's runtime is 16297.69 seconds ($\approx 4.5$ hours), compared to VI's 152.34 seconds ($\approx 2.5$ minutes) and the mixed logit's 414.48 seconds ($\approx 7$ minutes). In comparison to the baseline setting this corresponds to an increment of runtime of more than 100\% for MCMC, 50\% for VI and 171\% for the mixed logit model.

\begin{figure}[H] 
	\begin{center}
		
		\includegraphics[width=1\textwidth]{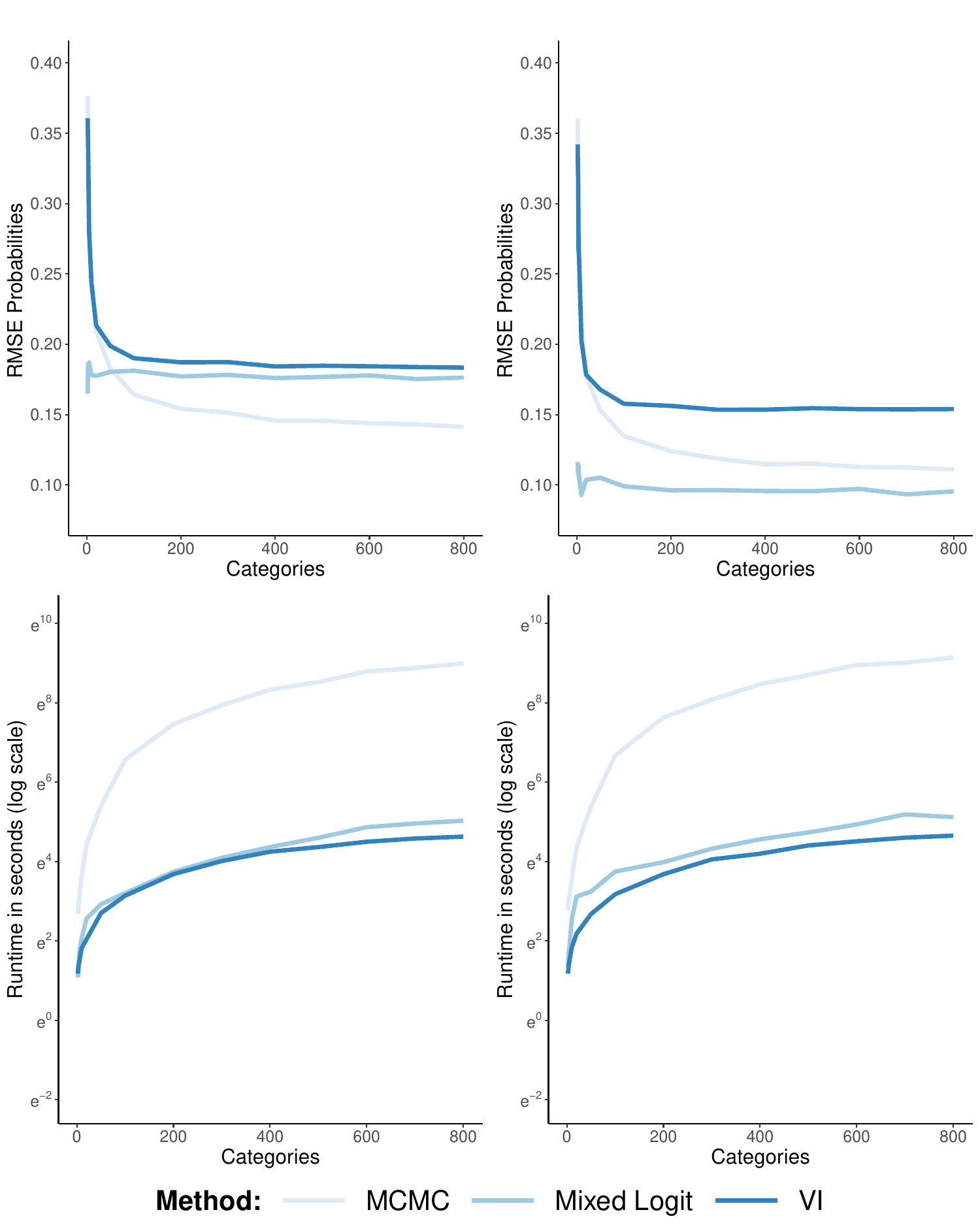}  
	\end{center}
	\vspace{-1ex}
	\caption{Comparison of $\text{RMSE}_{avg}$ and runtime of baseline setting (left column) and setting with  lower heterogeneity (right column). Both with $K = 3$, $I=20$ and $J_c=5$.}
	\label{abb: results_lowHet_plt}	
\end{figure}

In figure \ref{abb: results_lowHet_plt}, the right column presents the outcomes of the third setting, which assumes reduced heterogeneity in the individual and product specific latent variables. For comparative purposes, the results from the baseline setting, characterized by higher heterogeneity, are also displayed in the left column.
When analysing the runtime in the context of reduced heterogeneity, it is evident that this adjustment does not have a distinct impact on the computational duration.
The only notable difference is that the runtimes of VI and the mixed logit model are not as close as in the baseline setting.
The RMSEs for all methods exhibit significant reductions compared to the baseline. Specifically, VI's predictive accuracy narrows to an RMSE of 0.15, while MCMC achieves an RMSE of 0.11. Most notably, the mixed logit model experiences the greatest improvement, with its RMSE dropping to 0.09, a substantial improvement from its initial 0.17 in the baseline scenario.

\begin{table}[H]
	\centering
	\caption{Average runtime in seconds for estimation with $I=20$ and $I=40$. In both settings with $K=3$ and $J_c=5$. Rounded to two decimal places.}
	\label{tab:runtime_i40}
	\begin{tabular}{rrrrrrr}  
		\toprule 
		&  \multicolumn{3}{c}{$I = 20$}& \multicolumn{3}{c}{$I = 40$}  \\ \addlinespace
		C& MCMC & VI & Mixed logit & MCMC & VI & Mixed logit \\   
		\midrule
		2 & 14.48 & 3.24 & 2.94 & 21.74 & 4.87 & 5.06 \\   
		3& 17.13 & 3.79 & 3.36 & 25.36 & 5.11 & 6.98 \\   
		4& 20.60 & 4.11 & 4.61 & 30.78 & 5.61 & 8.99 \\   
		5& 22.51 & 4.34 & 5.47 & 35.06 & 5.63 & 9.43 \\   
		10 & 38.12 & 6.14 & 7.90 & 66.07 & 7.60 & 14.41 \\   
		20& 84.29 & 7.70 & 12.96 & 149.93 & 10.32 & 20.61 \\   
		50& 222.51 & 14.90 & 18.60 & 550.20 & 18.55 & 29.95 \\   
		100& 712.71 & 23.11 & 24.64 & 1795.19 & 32.17 & 57.42 \\   
		200& 1739.82 & 39.78 & 42.09 & 3837.22 & 54.75 & 96.27 \\   
		300& 2804.08 & 55.22 & 59.98 & 6411.43 & 76.59 & 182.59 \\   
		400& 4145.13 & 69.93 & 78.57 & 9168.08 & 98.10 & 217.00 \\   
		500& 5031.88 & 78.53 & 99.27 & 11158.20 & 107.73 & 280.49 \\   
		600& 6549.15 & 89.85 & 129.71 & 13963.86 & 126.20 & 337.53 \\   
		700& 7163.05 & 97.48 & 142.35 & 15178.64 & 134.98 & 374.15 \\   
		800& 8035.25 & 102.25 & 152.78 & 16297.69 & 152.34 & 414.48 \\    
		\bottomrule
	\end{tabular}
\end{table}

\subsection{Discussion}

The results of the simulation study raise several points of consideration, particularly regarding the fact that the average RMSEs do not converge to zero as the number of categories increases. Although the DGP follows the same latent factorization structure as the estimated model, this does not necessarily imply vanishing prediction errors. One possible explanation is the limited amount of information available for estimating product-specific latent factors. While additional categories provide more information about the shared individual-specific factors, they also introduce new product-specific parameters, each of which is informed by a limited number of individuals. Therefore, increasing the number of categories does not necessarily eliminate prediction uncertainty. The results from the second setting, where the number of individuals is increased, further suggest that additional household observations may improve predictive accuracy. However, increasing the number of individuals also raises the computational demands of the estimation methods.

Another limitation concerns the comparison with the mixed logit benchmark. Unlike the factorization model, the marginal mixed logit does not jointly account for repeated choices by the same individual. Furthermore, the benchmark assumes a single individual-specific price coefficient, whereas the DGP allows for product-specific price heterogeneity. Since the simulated data are generated according to the factorization model, the comparison inherently favors its specification. Consequently, the relatively constant predictive performance of the mixed logit model may partly reflect these differences rather than a general inability of mixed logit models to benefit from additional categories. Moreover, the third simulation setting changes not only the heterogeneity of the latent factors but also their means and the resulting utility scale. Therefore, improvements in predictive accuracy cannot be attributed exclusively to reduced heterogeneity.

Regarding the Bayesian estimation methods, one possibility to improve the predictive accuracy of VI is to use the \textit{fullrank} method in Stan, which allows for correlations between the parameters of the variational approximation instead of assuming a factorized Gaussian distribution \parencite{standev2023reference}. However, this approach is computationally more demanding and could reduce the runtime advantage of VI. For MCMC, increasing the maximum tree depth could improve sampling efficiency if the current limit restricts the exploration of the posterior distribution. Whether such adjustments are beneficial would need to be assessed using convergence diagnostics, such as effective sample sizes, divergences and tree-depth saturation. Therefore, the observed differences in predictive accuracy and runtime should also be interpreted in light of the respective estimation settings.

A further limitation of the simulation study is that choice probabilities are estimated conditional on purchasing a product from a given category. Consequently, the decision whether to purchase from a category at all is not explicitly modeled. This limitation could be addressed by a nested or stage-wise approach that includes an outside option, as implemented by \textcite{Donnelly2021counterfactual} and \textcite{wan2017modeling}.

Finally, the DGP considers only product prices as observed explanatory variables and does not explicitly account for substitution or complementarity across categories. These restrictive assumptions, together with the simplified mixed logit benchmark, limit the generalizability of the findings. Nevertheless, the simulation study provides insights into the predictive performance and computational requirements of MCMC and VI for latent factorization models under different dimensional settings.

\section{Real Data Application} \label{sec: real_data}

The objective of this section is to demonstrate the application of the models and methods to a large scale real-world dataset. For this application, the more complex version of the latent factorization model, defined in section \ref{sec: latent fac model}, is used
\begin{align}
	U_{ij_cts} &= \beta_i' \theta_{j_c}  - \gamma_i' \lambda_{j_c} \ p_{j_cts} + \epsilon_{ij_cts},
\end{align}
where $p_{j_cts}$ is the observed price of product $j$ in category $c$, week $t$ and store $s$. As stated before, the error term $\epsilon_{ij_cts}$ is assumed to be i.i.d. extreme value type 1 distributed.

The mixed logit model from section \ref{sec: MixedLogit} is augmented by a product specific intercept: 
\begin{align}
    U_{ij_cts} = \alpha_{j_c} + \beta_i \ p_{j_cts} + \epsilon_{ij_cts}.
\end{align}
The individual price sensitivity $\beta_i$ is assumed to be a random coefficient. 
Again, all choice probabilities are estimated conditional on the decision to buy on product from a given category.

\subsection{Data}

The models are estimated using scanner panel datasets provided by \textcite{Nielsen1988data}.\footnote{The documentation and raw data files are provided \href{https://www.chicagobooth.edu/research/kilts/research-data/erim}{online} from the Kilts Center for Marketing at the Chicago Booth School of Business.} Scanner panel data consists of longitudinal information on household purchasing, captured by electronic scanners in supermarkets. The available purchase data covers 10 product categories such as brownies, dinners, ketchup, margarine, dry detergents, canned soup, yogurt, sugar ,toilet tissue and canned tuna.

The data sets include households from two mid-size cities in the USA, namely Sioux Falls in South Dakota and Springfield in Missouri. In each of the markets 2500 households received magnetic ID cards to be presented at the checkout counter in 42 participating stores. The panels cover 186 weeks from February 1985 to August 1988. During this period, a total of 1.281.268 purchases are recorded, accounting for 80 \% of the grocery and drug retail sales in the markets \parencite{Nielsen1988data}. The purchase history files contain a range of variables, including market ID, week, day of the week, household ID, the \textit{universal product code} (UPC) of the bought product, store ID, total units purchased, and price. Additionally, various marketing variables such as coupons, advertising code, advertising type, and aisle display are documented, though they are not relevant to the current application. 

Not all categories are monitored throughout the entire duration. Consequently, the observation period is restricted to 75 weeks from January 1986 to the end of May 1987. Furthermore, each household's choice set is limited to the overall top 10 most frequently purchased products within each category. Because not every product is acquired across all combinations of categories, stores, and weeks, missing price values are imputed using predictions of the following regression:
\begin{align}
	p_{jcst} = \beta_0 + \beta_1 \cdot mean(p_{jc}) + week_t + store_s + \epsilon_{jcst}.
\end{align} 
\( p_{jcst} \) denotes the observed price per unit for product \( j \) in category \( c \) at store \( s \) during week \( t \), while \( \text{mean}(p_{jc}) \) represents the average price of product \( j \) in category \( c \) across all weeks and stores. $week_t$ and $store_s$ control for week and store fixed effects. The results of this regression can be found in appendix \ref{sec: App_B}. Note that prices are assumed to not change during the week.
Table \ref{tab: summary_data} provides a overview of the observations, distinct households and both mean and standard deviation of the prices at the category level.
Surprisingly, in five categories, the count of unique households exceeds the total of 5,000 households that were issued magnetic ID cards. One potential explanation is that some panelists might have exited the panel over the observation period, leading to the issuance of new ID cards. However, the available documentation does not provide any information on this matter.

\begin{table}[H]
	\centering
	\caption{Summary statistics of the cleaned data per category rounded to two decimal places.}
	\label{tab: summary_data}
	\begin{tabular}{lcccc}  
		\toprule 
		category &observations & households & mean price & sd price \\   
		\midrule 
		brownies & 7476 & 2683 & 1.50 & 0.38 \\   
		dinners & 4037 & 1722 & 1.54 & 0.48 \\   
		ketchup & 28916 & 5434 & 1.20 & 0.44 \\   
		margarine & 81915 & 6108 & 0.63 & 0.36 \\   
		dry detergents & 8716 & 1575 & 1.89 & 0.68 \\   
		canned soup & 51266 & 2711 & 0.41 & 0.14 \\   
		yogurt & 8993 & 1183 & 0.57 & 0.09 \\   
		sugar & 46849 & 5889 & 1.19 & 0.36 \\   
		toilet tissue & 72711 & 6138 & 0.92 & 0.21 \\   
		canned tuna & 61505 & 6016 & 0.64 & 0.28 \\ 
		\bottomrule
	\end{tabular}
\end{table}
To reduce computational demands, only the top 1000 households that shop most frequently are included. This results in a dataset with 148057 observations.

\subsection{Results}
For this application, the dataset is split randomly into three non-overlapping subsets: 60\% is allocated to the training set, while the validation and test sets each contain 20\% of the data.

In contrast to the prior simulation study where true choice probabilities for all alternatives were observable, in this context, only the purchase choices made by a household can be monitored. As a result, the predictive accuracy is assessed based on correctly predicted choices rather than choice probabilities. The RMSE of the mixed logit model is defined as
\begin{align}
	RMSE &= \sqrt{\sum_{n=1}^{N \cdot J_c} \frac{(\hat{y}_{n,j_c} - y^{\textit{obs}}_{n,j_c})^2}{N \cdot J_c}},
\end{align}
where $N$ denotes the number of observations in the validation or test dataset. $y_{n,j_c}^{obs}$ represents the observed product choice of the household in the $n$-th observation of the validation or test set 
\begin{align}
	y_{n,j_c}^{obs} = \begin{cases}
		1, \text{if household in the $n$-th observation buys product $j_c$}\\
		0, \text{otherwise}.
	\end{cases}
\end{align}
$\hat{y}_{n,j_c}$ is defined analogously for the predicted product choice of the household in the $n$-th observation.

Like in the previous simulation study, the RMSE of the latent factorization model is defined as the average RMSE over all samples $s = 1, \dots, S$ with $S=4000$:
\begin{align}
	RMSE = &\frac{1}{S} \sum_{s=1}^{S} RMSE_s \\
	& \text{with  } RMSE_s = \sqrt{\sum_{n=1}^{N \cdot J_c} \frac{(\hat{y}_{s,n,j_c} - y^{\textit{obs}}_{n,j_c})^2}{N \cdot J_c}},
\end{align}
where $\hat{y}_{s,n,j_c}$ is the predicted product choice of the household in the $n$-th observation of the validation or test dataset based on the $s$-th sample of posterior parameters.

The simulation study results indicate that MCMC does not scale well with an increasing number of individuals. Given the runtime constraints on the Hilbert cluster, MCMC failed to converge within a reasonable time frame. Hence, all presented results for the latent factorization model are based on estimates from VI.

Table \ref{tab: res_hypers_real} displays the outcomes from the hyperparameter search for $K$ conducted on the validation dataset. It becomes evident that a setting $K=10$ yields the lowest RMSE on the validation data. 
\begin{table}[H]
	\centering
	\caption{Results of hyperparameter search for $K$ performed with VI on 60\% training data and 20\% validation data for the 1000 most frequently shopping households.}
	\label{tab: res_hypers_real}
	\begin{tabular}{rcc}  
		\toprule 
		$K$ & runtime (in sec) & RMSE  \\   
		\midrule
		5 & 1165.58 & 0.800583 \\   \addlinespace
		\textbf{10} & \textbf{1726.26} & \textbf{0.791531} \\   \addlinespace
		15 & 2223.51 & 0.792237 \\   \addlinespace
		20 & 2561.81 & 0.793903 \\   \addlinespace
		25 & 2989.94 & 0.795423 \\   \addlinespace
		30 & 3295.88& 0.797239 \\    
		\bottomrule
	\end{tabular}
\end{table}


With 10-dimensional latent variables the resulting calculated RMSE on the held-out test dataset is 0.790812 and the computation process took 2040.16 seconds ($\approx 34$ minutes).
The mixed logit model required substantially more computational time than VI, taking nearly 5 times longer with a runtime of 9890.52 seconds ( $\approx$165 minutes). Furthermore, it delivered a higher RMSE of 0.848709 on the test dataset in comparison to VI.

		

\subsection{Discussion}

The results of the empirical application should be interpreted with caution, particularly regarding the evaluation of predictive accuracy. In contrast to the simulation study, the true choice probabilities are unknown, so predictive performance is evaluated using observed product choices. However, the reported RMSE values exceed the theoretical maximum implied by the stated one-hot prediction measure with ten alternatives. This indicates a discrepancy between the reported evaluation metric and its calculation that requires further investigation. Moreover, the predictive performance of VI is evaluated by averaging RMSEs across posterior draws, whereas the mixed logit model is evaluated using point estimates. Since this approach incorporates posterior parameter uncertainty into the VI prediction error, the comparison is not entirely equivalent. Evaluating posterior-mean choice probabilities using a common scoring rule would provide a more consistent comparison between the models.

Further limitations arise from the assumptions made in the model construction and estimation. Although the factorization model captures persistent household-specific preferences through shared latent factors, it does not explicitly account for temporal dynamics, such as state dependence or purchasing inertia. Furthermore, the mixed logit benchmark is estimated using a marginal likelihood that does not jointly account for repeated choices by the same household. Consequently, differences in predictive performance may partly reflect the models' different treatment of household purchase histories rather than differences in their general predictive capabilities. The selection of the 1000 most frequently shopping households may additionally introduce selection bias, limiting the representativeness of the results.

The empirical application is also subject to limitations arising from data preprocessing. In particular, the estimation requires prices for all products included in the choice sets across stores and weeks. Since these prices are not always observed, missing values are imputed using a regression model. However, the absence of an observed price does not necessarily imply that a product was available for purchase. Consequently, the constructed choice sets may contain unavailable alternatives, potentially affecting estimated choice probabilities. Furthermore, the use of product-level average prices and imputation parameters calculated from the full dataset may introduce information from validation or test observations into the training data. This potential data leakage could affect the reliability of the out-of-sample evaluation. The restriction to the ten most frequently purchased products per category and the treatment of multiple purchases within a category impose additional limitations on the representation of actual consumer choices.

Another concern relates to the identification of households across product categories. As discussed in Section 5.1, the number of distinct household identifiers exceeds the initially reported panel size in several categories. Without further verification of the household identifiers, potential inconsistencies in linking purchase histories cannot be ruled out. This is particularly relevant for the factorization model, which relies on shared household-specific latent factors across categories.

Although the models applied to the empirical data are more complex than those used in the simulation study, they remain simplified representations of consumer purchasing behavior. The mixed logit model could potentially benefit from incorporating demographic characteristics, time-fixed effects or more flexible random coefficients. Similarly, additional household- and product-specific observable characteristics could improve the latent factorization model. However, the empirical application is constrained by the information available in the scanner panel from \textcite{Nielsen1988data}. Furthermore, regional differences between Springfield and Sioux Falls are not explicitly modeled, as observations from both markets are combined in the training, validation and test datasets. Although the two regions exhibit demographic similarities \parencite{Nielsen1988data}, differences in local market conditions may still affect consumer preferences and purchasing behavior.

Finally, the computational requirements of MCMC remain an important limitation of the empirical application. Within the available computational resources, no sufficiently converged MCMC estimates were obtained for the large-scale dataset. However, this does not necessarily imply that MCMC is generally infeasible for such applications, as computational performance depends on the model implementation, sampling configuration and available hardware. Future research could investigate more efficient parameterizations, alternative sampling strategies and improved computational infrastructure. In particular, GPU-based computation might offer opportunities to reduce computational costs for suitable implementations \parencite{Owens2008GPU}.
\section{Conclusion} \label{sec: conclusion}

This work examined the application of Bayesian machine learning methods to large-scale demand estimation, focusing on MCMC and VI. Both approaches approximate complex posterior distributions but differ fundamentally in their computational procedures.

MCMC methods approximate posterior distributions by constructing Markov chains whose invariant distribution corresponds to the target posterior. The MH algorithm achieves this through a proposal distribution and an acceptance rule that ensures detailed balance. However, its efficiency depends strongly on the choice of the proposal distribution, particularly in high-dimensional parameter spaces. HMC addresses some of these limitations by introducing auxiliary momentum variables and using Hamiltonian dynamics to explore the parameter space more efficiently. Since the performance of HMC depends on the step size and trajectory length, the NUTS Sampler provides an adaptive extension that reduces the need for manual tuning.


In contrast, VI methods transform the challenge of approximating the posterior distribution to an optimization problem.
The objective is the identification of a specific member within the variational family that minimizes the KL-divergence between the variational and the desired posterior distributions.
The CAVI algorithm solves this problem by iteratively optimizing individual variational factors while keeping the remaining factors fixed. This iterative approach is computationally demanding, limiting its application for large datasets.
ADVI presents an innovative approach to variational inference by transforming the latent variable space, expressing gradients as expectations, utilizing a standard Gaussian for re-parametrization, and employing stochastic optimization techniques. This provides a flexible alternative to computationally intensive posterior sampling.

In the simulation study, a latent factorization model was estimated using MCMC (NUTS) and ADVI and compared with a mixed logit benchmark. To investigate the effects of increasing dimensionality, fifteen settings with between 2 and 800 product categories were considered. Additionally, two alternative settings were examined, increasing the number of individuals and modifying the distributions of individual- and product-specific latent factors.

The results indicate that the latent factorization model benefits from information sharing across categories, leading to improvements in predictive accuracy as the number of observed categories increases. Within the factorization model, MCMC generally achieves higher predictive accuracy than VI, but at substantially greater computational cost, particularly as the number of individuals increases. VI provides a computationally more efficient alternative, although this advantage comes with a loss in predictive accuracy in the investigated simulation settings. The relative performance of the mixed logit benchmark depends on the simulation setting, with particularly strong results under the modified lower-heterogeneity scenario

However, these findings must be interpreted in light of the model specifications. The simulation data are generated according to the factorization model, while the mixed logit benchmark is estimated using a marginal likelihood that does not jointly account for repeated choices by the same individual. Furthermore, the mixed logit assumes a single individual-specific price coefficient, whereas the DGP allows for product-specific price heterogeneity. Consequently, the relatively constant predictive performance of the mixed logit benchmark cannot be interpreted as a general inability of mixed logit models to benefit from additional categories.

In the application with a large scale supermarket purchase dataset, the high computational effort of MCMC rendering it infeasible for this large scale data. VI not only surpasses the mixed logit model in terms of computational efficiency but also exhibits superior predictive accuracy.

An important limitation of both the simulation and empirical application is that choice probabilities are estimated conditional on purchasing a product from a given category. To maintain more generalized results, future research should address this limitation. For example, by incorporating nested or stage wise models like in the work of \textcite{Donnelly2021counterfactual} or \textcite{wan2017modeling}. 
Another direction for future research is the enhancement of the computational performance of the models. Given the computational demands associated with these models, transitioning from CPU-based computation to GPU-based computation could offer significant improvements in processing speed and efficiency. Leveraging the parallel processing capabilities of GPUs may facilitate more timely results, particularly for large-scale applications and complex datasets \parencite{Owens2008GPU}.

Finally, it should be noted that it is possible to combine MCMC and VI. In this way, additional computing time can be traded off against increased accuracy \parencite{Salimans_et_al_2017combine}. Overall, the findings demonstrate the potential of Bayesian latent factorization methods for multi-category demand estimation while highlighting the importance of balancing predictive accuracy, model flexibility and computational feasibility.


\onecolumn 
\singlespacing 
\newpage

\newpage
\addcontentsline{toc}{section}{References}

\printbibliography[title=References]

\newpage

\addcontentsline{toc}{section}{Appendix}

\appendix
\renewcommand{\thesubsection}{\Alph{subsection}}
\section*{Appendix}
\subsection{Derivation of CAVI} \label{sec: App_A}
Show that the optimal variational factor $q_j^*(\theta_j)$ for a given latent variable $\theta_j$ in the mean-field variational familiy is given by:
\begin{align*}
	q_j^*(\theta_j) &= \underset{q_j}{argmax} \ ELBO(q_j),\\
	&=  \frac{exp\big(\mathbb{E}_{q_{i\neq j}}\big[log \ p(X, \theta_{i\neq j}, \theta_j)\big]\big)}{\int exp\big(\mathbb{E}_{q_{i\neq j}}\big[log \ p(X, \theta_{i\neq j}, \theta_j)\big]\big) d\theta_j} \propto exp\big(\mathbb{E}_{q_{i\neq j}}\big[log \ p(X, \theta_{i\neq j}, \theta_j)\big]\big).
\end{align*}
Start with the objective function from equation \eqref{eq: ELBO_qj}:
\begin{align*}
	ELBO(q_j) &= \int \big[q_j \prod_{i \neq j}^{K} q_i\big] \big[log \ p(X, \theta) - (log \ q_j + \sum_{i\neq j}^{K} log \ q_i )\big] d\theta, \\
	&= \int q_j \int \big[\prod_{i \neq j}^K q_i\big] \big[log \ p(X, \theta) - (log \ q_j + \sum_{i\neq j}^K log \ q_i)\big] d\theta_{i \neq j} \ d\theta_j, \\
	&= \underbrace{\int q_j \int \big[\prod_{i \neq j}^K q_i\big] \big[log \ p(X, \theta)\big] d\theta_{i \neq j} \ d\theta_j}_{\textbf{1.}} \\
	& \ \ \ \ \ \ \ \ \ \ \ - \underbrace{\int q_j \int \big[\prod_{i \neq j}^K q_i\big] \big[log \ q_j + \sum_{i\neq j}^Klog \ q_i)\big] d\theta_{i \neq j} \ d\theta_j}_{\textbf{2.}}
\end{align*}

In the first term one can rewrite $p(X, \theta) = p(X, \theta_{i \neq j}, \theta_j)$, so the other variables are grouped apart from $\theta_j$: 
\begin{align*}
		&\int q_j \int \big[\prod_{i \neq j}^K q_i\big] \big[log \ p(X, \theta)\big] d\theta_{i \neq j} \ d\theta_j, \\
		=& \int q_j \  \mathbb{E}_{q_{i \neq j}} \big[log \ p(X, \theta_{i \neq j}, \theta_j)\big] d\theta_j, \\
		=& \int log \ \tilde{p}(X, \theta_j) - S \ d\theta_j, \\
		=& \int log  \ \tilde{p}(X, \theta_j) \ d\theta_j - \int S \ d\theta_j, \\
		=& \int log  \ \tilde{p}(X, \theta_j) \ d\theta_j + S', \numberthis \label{eq: a_first}
\end{align*}
where $S$ and $S'$ are constants and $\tilde{p}(X, \theta_j)$ is defined as 
\begin{align}
	log \ \tilde{p}(X, \theta_j) = \mathbb{E}_{q_{i \neq j}} \big[log \ p(X, \theta_{i \neq j}, \theta_j)\big] + S. \label{eq: a_def_p}
\end{align}
The second term can be rewritten in the following way:
\begin{align*}
	& \int q_j \int \big[\prod_{i \neq j}^K q_i\big] \big[log \ q_j + \sum_{i\neq j}^K log \ q_i)\big] d\theta_{i \neq j} \ d\theta_j, \\
	=& \int q_j \int \big[\prod_{i \neq j}^K q_i\big] \ log \ q_j \ d\theta_{i \neq j} \ d\theta_j + \int q_j \underbrace{\int \big[\prod_{i \neq j}^K q_i\big] \sum_{i\neq j}^K log \ q_i \ d\theta_{i \neq j}}_{= \widehat{S}} \ d\theta_j, \\
	=&  \int q_j \int \big[\prod_{i \neq j}^K q_i\big] \ log \ q_j \ d\theta_{i \neq j} \ d\theta_j + \widehat{S} \underbrace{\int q_j \ d\theta_j}_{= 1}, \\
	=& \int q_j \int \big[\prod_{i \neq j}^K q_i\big] \ log \ q_j \ d\theta_{i \neq j} \ d\theta_j + \widehat{S}, \\
	=& \int q_j \ log \ q_j \underbrace{\int \prod_{i \neq j}^K q_i \ d\theta_{i \neq j} }_{ = 1} \ d\theta_j + \widehat{S}, \\
	=& \int q_j \ log\ q_j \ d\theta_j + \widehat{S}, \numberthis \label{eq: a_sec}
\end{align*}
where $\widehat{S}$ is a constant with respect to $q_j$. With equations \eqref{eq: a_first} and \eqref{eq: a_sec} the ELBO is the negative KL-divergence of $\tilde{p}(X, \theta_j)$ and $q_j(\theta_j)$
\begin{align*}
	ELBO(q_j) &= \int log  \ \tilde{p}(X, \theta_j) \ d\theta_j + \int q_j \ log\ q_j \ d\theta_j + C, \\
	&= - D_{KL} \big(q_j(\theta_j) \ || \ \tilde{p}(X, \theta_j)\big).
\end{align*}
$S'$ and $\widehat{S}$ are absorbed into constant $C$. Setting $q_j^*(\theta_j) = \tilde{p}(X, \theta_j)$ maximizes the ELBO, because it causes KL-divergence to be zero. With the definition in equation \eqref{eq: a_def_p}  the optimal choice for $q_j$ is
\begin{align*}
	q_j^*(\theta_j) &= exp\big(\mathbb{E}_{q_{i \neq j}} \big[log \ p(X, \theta_{i \neq j}, \theta_j)\big] + S\big), \\
	&= exp\big(\mathbb{E}_{q_{i \neq j}} \big[log \ p(X, \theta_{i \neq j}, \theta_j)\big] \big) \cdot \underbrace{exp(S)}_{\text{normalization constant}}, \\
	&=  \frac{exp\big(\mathbb{E}_{q_{i\neq j}}\big[log \ p(X, \theta_{i\neq j}, \theta_j)\big]\big)}{\int exp\big(\mathbb{E}_{q_{i\neq j}}\big[log \ p(X, \theta_{i\neq j}, \theta_j)\big]\big) d\theta_j}, \\
	& \\
	&\propto exp\big(\mathbb{E}_{q_{i\neq j}}\big[log \ p(X, \theta_{i\neq j}, \theta_j)\big]\big).
\end{align*}

\subsection{Regression result for price interpolation} \label{sec: App_B}

\begin{table}[H] \centering  
	 \caption{Regression table for price regression with week and store fixed effects.}  
	 \label{tab: reg_price} 
	 \begin{tabular}{@{\extracolsep{5pt}}lc} \\[-1.8ex]\hline \hline \\[-1.8ex]  & \multicolumn{1}{c}{\textit{Dependent variable:}} \\ \cline{2-2} \\[-1.8ex] & $p_{jcst}$\\ \hline \\[-1.8ex]  Mean($p_{jc}$)& 0.997$^{***}$ (0.001) \\   Constant & 0.059$^{***}$ (0.005) \\  \hline \\[-1.8ex] Week & Yes \\ Store & Yes \\ \hline \\[-1.8ex] Observations & 88,431 \\ R$^{2}$ & 0.939 \\ Adjusted R$^{2}$ & 0.939 \\ Residual Std. Error & 0.136 (df = 88314) \\ F Statistic & 11,799.850$^{***}$ (df = 116; 88314) \\ \hline \hline \\[-1.8ex] \textit{Note:}  & \multicolumn{1}{r}{$^{*}$p$<$0.1; $^{**}$p$<$0.05; $^{***}$p$<$0.01} \\ 
	  \end{tabular} 
\end{table}

\end{document}